\documentclass[10pt]{article}
\usepackage{graphicx}
\usepackage{amsmath}
\usepackage{amssymb}
\usepackage{caption2}
\begin{document}
\begin{center}
{\large {\bf \sc{  Landau equation and QCD sum rules for the tetraquark molecular states
  }}} \\[2mm]
Zhi-Gang  Wang \footnote{E-mail: zgwang@aliyun.com.  }  \\
 Department of Physics, North China Electric Power University, Baoding 071003, P. R. China
\end{center}

\begin{abstract}
The quarks and gluons are confined objects, they cannot be put on the mass-shell, it is questionable  to  apply  the Landau equation to study the Feynman diagrams in the QCD sum rules. Furthermore,  we carry out the operator product expansion in the deep Euclidean region $p^2\to -\infty$, where the Landau singularities cannot exist. The Landau equation servers  as a kinematical equation in the momentum space, and is independent on the factorizable and nonfactorizable properties of the Feynman diagrams in the color space.  The meson-meson scattering state and tetraquark molecular state both have four valence quarks, which form two color-neutral clusters, we cannot distinguish the contributions  based on the two color-neutral clusters in the factorizable Feynman diagrams.
Lucha, Melikhov and Sazdjian assert that the contributions  at the order $\mathcal{O}(\alpha_s^k)$ with $k\leq1$ in the operator product expansion, which are factorizable in the color space, are exactly  canceled out   by the meson-meson scattering states at the hadron side, the tetraquark molecular states begin to receive contributions at the order $\mathcal{O}(\alpha_s^2)$. Such an assertion is questionable, we refute the assertion in details, and  choose an axialvector current and a tensor current  to examine the outcome of the assertion. After detailed  analysis, we observe that the meson-meson scattering states cannot saturate the QCD sum rules,  while
the tetraquark molecular states can saturate the QCD sum rules.
The Landau equation is of no use  to study  the Feynman diagrams in the QCD sum rules for the tetraquark molecular states,  the   tetraquark molecular states begin to receive contributions at the order $\mathcal{O}(\alpha_s^0/\alpha_s^1)$ rather than at the order $\mathcal{O}(\alpha_s^2)$.
\end{abstract}

 PACS number: 12.39.Mk, 12.38.Lg

Key words: Molecular   state, QCD sum rules

\section{Introduction}

In 2003, the  Belle collaboration   observed   a narrow charmonium-like structure $X(3872)$ in the $\pi^+ \pi^- J/\psi$ invariant mass spectrum in the exclusive $B$-decays \cite{X3872-2003}, which cannot be accommodated in the traditional or normal  quark-antiquark model. Thereafter, more than twenty   charmonium-like exotic  states were observed  by the  BaBar, Belle, BESIII, CDF, CMS, D0, LHCb collaborations \cite{PDG},  some exotic  states are still needed confirmation and their quantum numbers have not been established yet.
There have seen several possible interpretations for those $X$, $Y$ and $Z$ states, such as the tetraquark states, tetraquark (or hadronic) molecular states, dynamically generated resonances,  hadroquarkonium,
kinematical effects, cusp effects, etc \cite{HXChen-review-1601,MNielsen-review-1812}.

Among those possible interpretations, the tetraquark states and tetraquark  molecular states are outstanding and attract much attention as the exotic $X$, $Y$ and $Z$ states lie near the  thresholds of two charmed mesons.
In 2006, R. D. Matheus et al assigned  the $X(3872) $ to be  the $J^{PC}=1^{++}$ diquark-antidiquark type  tetraquark state, and studied  its mass with the QCD sum rules \cite{Narison-3872}. It is the first time to apply the QCD sum rules to study the exotic $X$, $Y$ and $Z$ states.  Thereafter  the QCD sum rules become a powerful theoretical approach in studying the masses and widths of the exotic $X$, $Y$ and $Z$ states, irrespective of assigning them as the hidden-charm (or hidden-bottom) tetraquark states or tetraquark (or hadronic) molecular states, and have  given many successful descriptions of the hadron properties \cite{MNielsen-review-1812,Narison-3872,QCDSR-4-quark-mass,WangHuangtao-PRD,Wang-tetra-formula,WangHuang-2014-NPA,
QCDSR-4-quark-width,WangZG-4-quark-mole,WangZG-CPC-Y4390}.
In the QCD sum rules for the tetraquark states and tetraquark molecular states, we choose the diquark-antidiquark type currents or meson-meson type (more precisely,  the color-singlet-color-singlet type currents), respectively, they can be reformed into each other via Fierz rearrangements, for example,
\begin{eqnarray}\label{Fierz}
J_{\mu}&=&\frac{\varepsilon^{ijk}\varepsilon^{imn}}{\sqrt{2}}\Big\{u^{T}_jC\gamma_5 c_k \bar{d}_m\gamma_\mu C \bar{c}^{T}_n
-u^{T}_jC\gamma_\mu c_k\bar{d}_m\gamma_5 C \bar{c}^{T}_n \Big\} \, , \nonumber\\
&=&\frac{1}{2\sqrt{2}}\Big\{\,i\bar{c}i\gamma_5 c\,\bar{d}\gamma_\mu u-i\bar{c} \gamma_\mu c\,\bar{d}i\gamma_5 u+\bar{c} u\,\bar{d}\gamma_\mu\gamma_5 c
-\bar{c} \gamma_\mu \gamma_5u\,\bar{d}c  \nonumber\\
&&  - i\bar{c}\gamma^\nu\gamma_5c\, \bar{d}\sigma_{\mu\nu}u+i\bar{c}\sigma_{\mu\nu}c\, \bar{d}\gamma^\nu\gamma_5u
- i \bar{c}\sigma_{\mu\nu}\gamma_5u\,\bar{d}\gamma^\nu c+i\bar{c}\gamma^\nu u\, \bar{d}\sigma_{\mu\nu}\gamma_5c   \,\Big\} \, ,
\end{eqnarray}
where the $i$, $j$, $k$, $m$, $n$ are color indices.

In the correlation functions for the color-singlet-color-singlet type currents,   Lucha, Melikhov and Sazdjian assert that  the Feynman diagrams can be divided into or separated into  factorizable diagrams and nonfactorizable diagrams in the color space in the operator product expansion,
 the contributions  at the order $\mathcal{O}(\alpha_s^k)$ with $k\leq1$, which are factorizable in the color space, are exactly  canceled out    by the meson-meson scattering states at the hadron side,
the nonfactorizable diagrams, if have a Landau singularity, begin to make contributions  to the tetraquark (molecular) states,  the tetraquark (molecular) states begin to receive contributions at the order $\mathcal{O}(\alpha_s^2)$  (according to the Fierz rearrangements, see Eq.\eqref{Fierz})  \cite{Chu-Sheng-PRD}.

About ten years  before the work of Lucha, Melikhov and Sazdjian,  Lee and Kochelev studied  the two-pion  contributions in the QCD sum rules for the scalar meson $f_0(600)$ (or $f_0(500)$ named by the Particle Data Group now \cite{PDG}) as the tetraquark state, and observed that the contributions of the order $\mathcal{O}(\alpha_s^k)$  with $k\leq1$ cannot be canceled out by the two-pion  scattering states \cite{Lee-0702-PRD}.

In this article, we will examine the assertion of Lucha, Melikhov and Sazdjian
in details  and use two examples to illustrate that the Landau equation is of no use in the QCD sum rule for the tetraquark molecular states.

The article is arranged as follows: in Sect.2, we discuss  the usefulness of the Landau equation in
the QCD sum rules for the tetraquark molecular states;
in Sect.3, we obtain the QCD sum rules for
the meson-meson scattering states  and tetraquark molecular states  as an example; in Sect.4,
we present the numerical results and discussions; Sect.5 is reserved for our conclusion.

\section{Is Landau equation useful  in the QCD sum rules for the tetraquark molecular states?}

In the following, we write down  the two-point correlation function $\Pi_{\mu\nu}(p)$  in the QCD sum rules as an example,
\begin{eqnarray}
\Pi_{\mu\nu}(p)&=&i\int d^4x e^{ip \cdot x} \langle0|T\Big\{J_\mu(x)J_\nu^{\dagger}(0)\Big\}|0\rangle \, ,
\end{eqnarray}
where
\begin{eqnarray}
J_\mu(x)&=&\frac{1}{\sqrt{2}}\Big[\bar{u}(x)i\gamma_5 c(x)\bar{c}(x)\gamma_\mu d(x)+\bar{u}(x)\gamma_\mu c(x)\bar{c}(x)i\gamma_5 d(x) \Big]  \, .
\end{eqnarray}
The color-singlet-color-singlet type current $J_\mu(x)$ has the quantum numbers $J^{PC}=1^{+-}$, at the hadron side, the quantum field theory allows  non-vanishing couplings to the  $D\bar{D}^*+D^*\bar{D}$ scattering states or tetraquark molecular states with the  $J^{PC}=1^{+-}$.

\begin{figure}
 \centering
  \includegraphics[totalheight=4cm,width=5cm]{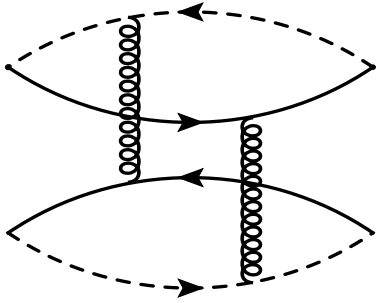}
 \caption{ The nonfactorizable  Feynman diagrams of the order $\mathcal{O}(\alpha_s^2)$ for the color-singlet-color-singlet  type currents, other diagrams obtained by interchanging of the heavy quark lines (dashed lines) and light quark lines (solid lines) are implied.}\label{meson-afs2}
\end{figure}

At the QCD side, when we carry out the operator product expansion, Lucha, Melikhov and Sazdjian assert  that the Feynman diagrams can be divided into or  separated  into factorizable diagrams and nonfactorizable diagrams, the Feynman diagrams of the orders $\mathcal{O}(\alpha_s^0)$ and $\mathcal{O}(\alpha_s^1)$ are factorizable,  the factorizable diagrams are exactly  canceled out by   the meson-meson scattering states, while the nonfactorizable Feynman diagrams, which are of the order $\mathcal{O}(\alpha_s^2)$, if  have a Landau singularity, begin to  make contributions  to the tetraquark (molecular) states,  the tetraquark (molecular) states begin to receive contributions at the order $\mathcal{O}(\alpha_s^2)$ \cite{Chu-Sheng-PRD},  see the Feynman diagrams shown Fig.\ref{meson-afs2}. In fact, such an assertion is questionable.

{\bf Firstly}, we cannot assert that the factorizable  Feynman diagrams in color space are exactly canceled out by the meson-meson scattering states, because the meson-meson scattering state and tetraquark molecular state both have four valence quarks, which can be divided into or  separated into two color-neutral clusters. We cannot distinguish which Feynman diagrams contribute to the  meson-meson scattering state or tetraquark molecular state based on the two color-neutral clusters.

{\bf Secondly}, the quarks and gluons are confined objects, they cannot be put on the mass-shell, it is questionable  to assert  that the Landau equation is applicable  in the nonperturbative  calculations dealing with  the quark-gluon bound states \cite{Landau}.

If we insist on applying  the Landau equation to study the Feynman diagrams in the QCD sum rules, we should choose the pole masses rather than the $\overline{MS}$ masses to warrant that  there  exists a mass pole which corresponds to the mass-shell in pure perturbative calculations, just like in the quantum  electrodynamics, where the electron, muon and tau can be put on the mass-shell.

According to the assertion of Lucha, Melikhov and Sazdjian, the tetraquark (molecular) states begin to receive contributions at the order $\mathcal{O}(\alpha_s^2)$ \cite{Chu-Sheng-PRD},
it is reasonable to take the pole masses $\hat{m}_Q$ as,
\begin{eqnarray}
\hat{m}_Q&=&m_Q(m_Q)\left[1+\frac{4}{3}\frac{\alpha_s(m_Q)}{\pi}+f\left(\frac{\alpha_s(m_Q)}{\pi}\right)^2+g\left(\frac{\alpha_s(m_Q)}{\pi}\right)^3\right]\, ,
\end{eqnarray}
to put the heavy quark lines on the mass-shell, the explicit expressions of the  coefficients $f$  and $g$ can be found in Refs.\cite{PDG,Three-loop-mass}. It is
straightforward to obtain
$ \hat{m}_b=m_b(m_b)\left(1 + 0.10 + 0.05 + 0.03\right)=4.78\pm0.06\,\rm{GeV}$ \cite{PDG}.

If the Landau equation is applicable in the QCD sum rules for the tetraquark states and tetraquark molecular states, it is certainly applicable in the QCD sum rules for
the traditional or normal charmonium and bottomonium states.
In the case of the $c$-quark, the pole mass $\hat{m}_c=1.67\pm0.07\,\rm{GeV}$ from the Particle Data Group \cite{PDG}, the Landau singularity appears at the $s$-channel  $\sqrt{s}=\sqrt{p^2}=2\hat{m}_c=3.34\pm0.14\,{\rm{GeV}}>m_{\eta_c}$ and $m_{J/\psi}$. While in the case of the $b$-quark, the pole mass $\hat{m}_b=4.78\pm0.06\,\rm{GeV}$ from the Particle Data Group \cite{PDG}, the Landau singularity appears at the $s$-channel $\sqrt{s}=\sqrt{p^2}=2\hat{m}_b=9.56\pm0.12\,{\rm{GeV}}>m_{\eta_b}$ and $m_{\Upsilon}$. It is odd or unreliable  that the  masses of the charmonium (bottomonium) states lie below the threshold $2\hat{m}_c$ ($2\hat{m}_b$) in the QCD sum rules for the $\eta_c$ and $J/\psi$ ($\eta_b$ and $\Upsilon$), as the integrals of the forms
\begin{eqnarray}
\int_{4\hat{m}_c^2}^{s_0} \delta\left(s-m_{\eta_c/J/\psi}^2 \right)\exp\left( -\frac{s}{T^2}\right) ds \, , \nonumber \\
\int_{4\hat{m}_b^2}^{s_0} \delta\left(s-m_{\eta_b/\Upsilon}^2 \right)\exp\left( -\frac{s}{T^2}\right) ds \, ,
\end{eqnarray}
at the hadron side are meaningless, where the $T^2$ is the Borel parameter. The tiny widths of the $\eta_c$, $J/\psi$, $\eta_b$ and $\Upsilon$  valuate the zero-width approximation, the hadronic spectral densities are of the form $\delta\left(s-m_{\eta_c/J/\psi/\eta_b/\Upsilon}^2 \right)$.

{\bf Thirdly}, the nonfactorizable Feynman diagrams which have the Landau singularities begin to appear at the order $\mathcal{O}(\alpha_s^0/\alpha_s^1)$ rather than at the order $\mathcal{O}(\alpha_s^2)$, and make contributions  to the tetraquark molecular states, if the assertion (the nonfactorizable Feynman diagrams which have Landau singularities make contributions to the tetraquark molecular states) of Lucha, Melikhov and Sazdjian is right.

\begin{figure}
 \centering
  \includegraphics[totalheight=4cm,width=5cm]{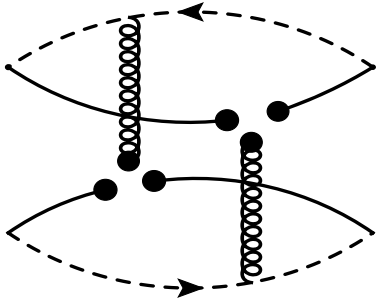}
 \caption{ The nonfactorizable   Feynman diagrams contribute to the vacuum condensates
$\langle \bar{q}g_s\sigma G q \rangle^2$ for the color-singlet-color-singlet  type currents, where the solid lines and dashed lines denote the light quarks and heavy quarks, respectively. }\label{meson-qqg-qqg}
\end{figure}

The nonperturbative contributions play an important role  and serve as a  hallmark for  the nonperturbative nature of the QCD sum rules,  the nonfactorizable contributions appear at the  order $\mathcal{O}(\alpha_s)$ due to the   operators  $\bar{q}g_sGq\bar{q}g_sGq$, which  come from the Feynman diagrams shown in Fig.\ref{meson-qqg-qqg}.
Such Feynman diagrams can be taken as annihilation diagrams, which play an important role in the tetraquark molecular states \cite{FKGuo1308}.
If we insist on applying the landau equation to study the Feynman diagrams shown in Fig.\ref{meson-qqg-qqg} and choose the  pole mass of the $c$-quark, we obtain a sub-leading Landau singularity at the $s$-channel $s=p^2=(\hat{m}_c+\hat{m}_c)^2$, which indicates that it contributes to the tetraquark molecular states.
From the operators  $\bar{q}g_sGq\bar{q}g_sGq$, we can obtain the vacuum condensate $\langle\bar{q}g_s\sigma Gq\rangle^2$,  where the $g_s^2=4\pi \alpha_s$ is absorbed into the vacuum condensate, so the Feynman  diagrams in  Fig.\ref{meson-qqg-qqg} can be counted as of the order $\mathcal{O}(\alpha_s^0)$. The  nonfactorizable Feynman diagrams appear at the order  $\mathcal{O}(\alpha_s^0)$
or $\mathcal{O}(\alpha_s^1)$ (based on how to account for the $g_s^2$ in the vacuum condensates), not at the order $\mathcal{O}(\alpha_s^2)$ asserted in Ref.\cite{Chu-Sheng-PRD}.

{\bf Fourthly}, the Landau equation servers  as a kinematical equation in the momentum space, and is independent on the factorizable and nonfactorizable properties of the Feynman diagrams in the color space. Without taking it for granted that the  factorizable Feynman diagrams in the  color space only make contributions to the two-meson scattering states,  the Landau equation cannot exclude the factorizable Feynman diagrams in the color space, those diagrams can also have the Landau singularities.

\begin{figure}
 \centering
  \includegraphics[totalheight=6cm,width=8cm]{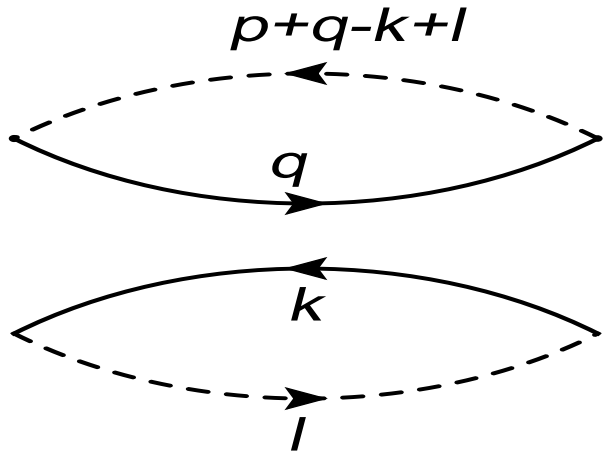}
 \caption{ The  Feynman diagrams  for the lowest order  contributions, where the solid lines and dashed lines represent  the light quarks and heavy quarks, respectively. }\label{Lowest-diagram}
\end{figure}

In the leading order, the factorizable Feynman diagrams shown in Fig.\ref{Lowest-diagram} can be divided into or  separated into two color-neutral clusters, each cluster corresponds to a trace both in the color space and in the Dirac spinor space. However, in the momentum space, they are nonfactorizable diagrams, the basic integrals are of the form,
\begin{eqnarray}\label{Basic-Integral}
\int d^4q d^4k d^4l \frac{1}{\left(p+q-k+l\right)^2-m_c^2}\frac{1}{q^2-m_q^2}\frac{1}{k^2-m_q^2}\frac{1}{l^2-m_c^2}\, .
\end{eqnarray}
If we choose the pole masses, there exists a  Landau singularity or an  $s$-channel singularity at $s=p^2=(\hat{m}_u+\hat{m}_d+\hat{m}_c+\hat{m}_c)^2$, which is just a signal of a four-quark intermediate state. We cannot assert that it is a signal of a meson-meson scattering state or a tetraquark molecular state, because the meson-meson scattering state and  tetraquark molecular state both have four valence quarks,  $q$, $\bar{q}$, $c$ and $\bar{c}$, which form  two color-neutral clusters.
The Landau singularity   is just a  kinematical singularity, not a dynamical singularity \cite{FKGuo-cusps}, it is useless in distinguishing  the contributions to the meson-meson scattering state and  tetraquark molecular state.  If we switch off the assertion that the
factorizable Feynman diagrams shown in Fig.\ref{Lowest-diagram} make contributions to the meson-meson scattering states alone, the $s$-channel singularity at $s=p^2=(\hat{m}_u+\hat{m}_d+\hat{m}_c+\hat{m}_c)^2$ supports that they also contribute to the tetraquark molecular states.

{\bf Fifthly},  only formal QCD sum rules for the tetraquark states or tetraquark molecular states are obtained based on the assertion of Lucha,  Melikhov and  Simula
in Ref.\cite{Chu-Sheng-PRD},
no feasible QCD sum rules with predictions can be confronted  to the experimental data are obtained up to now.

{\bf Sixthly},  in the QCD sum rules, we carry out the operator product expansion in  the deep Euclidean space, $-p^2 \to \infty$,  then obtain the physical spectral densities at the quark-gluon level  through dispersion relation \cite{SVZ79,Reinders85,ColangeloReview},
 \begin{eqnarray}
 \rho_{QCD}(s)&=&\frac{1}{\pi}{\rm Im}\,\Pi(s+i\epsilon)\mid_{\epsilon\to 0}\, ,
 \end{eqnarray}
 where the $\Pi(s)$ denotes the correlation functions. The Landau singularities require that the squared momentum $p^2=(\hat{m}_u+\hat{m}_d+\hat{m}_c+\hat{m}_c)^2$ in the Feynman diagrams, see Fig.\ref{Lowest-diagram} and Eq.\eqref{Basic-Integral}, it is questionable  to perform the operator product expansion.

\section{QCD sum rules with color-singlet-color-singlet type currents }

Now let us assume that the assertion of Lucha, Melikhov and Sazdjian is right, the tetraquark molecular states begin to receive contributions at the order $\mathcal{O}(\alpha_s^2)$, the contributions at the order $\mathcal{O}(\alpha_s^k)$ with $k\leq1$ are exactly canceled out by  the meson-meson scattering states.
We saturate the QCD sum rules with the meson-meson scattering  states and examine whether or not we can obtain feasible QCD sum rules.

In the following, we write down  the two-point correlation functions $\Pi_{\mu\nu}(p)$ and $\Pi_{\mu\nu\alpha\beta}(p)$ in the QCD sum rules,
\begin{eqnarray}
\Pi_{\mu\nu}(p)&=&i\int d^4x e^{ip \cdot x} \langle0|T\Big\{J_\mu(x)J_\nu^{\dagger}(0)\Big\}|0\rangle \, , \\
\Pi_{\mu\nu\alpha\beta}(p)&=&i\int d^4x e^{ip \cdot x} \langle0|T\left\{J_{\mu\nu}(x)J_{\alpha\beta}^{\dagger}(0)\right\}|0\rangle \, ,
\end{eqnarray}
where
\begin{eqnarray}
J_\mu(x)&=&\frac{1}{\sqrt{2}}\Big[\bar{u}(x)i\gamma_5 c(x)\bar{c}(x)\gamma_\mu d(x)+\bar{u}(x)\gamma_\mu c(x)\bar{c}(x)i\gamma_5 d(x) \Big]  \, ,  \\
J_{\mu\nu}(x)&=&\frac{1}{\sqrt{2}}\Big[\bar{s}(x)\gamma_\mu c(x) \bar{c}(x)\gamma_\nu\gamma_5  s(x)-\bar{s}(x)\gamma_\nu\gamma_5 c(x) \bar{c}(x)\gamma_\mu s(x)\Big]\, . \end{eqnarray}
The current $J_\mu(x)$ has the quantum numbers $J^{PC}=1^{+-}$, while the current $J_{\mu\nu}(x)$ has definite charge conjugation, the components $J_{0i}(x)$ and
  $J_{ij}(x)$ have positive-parity and negative-parity, respectively,  where the space indexes $i$, $j=1$, $2$, $3$.
  The charged current  $J_\mu(x)$ couples potentially to the  $D\bar{D}^*+D^*\bar{D}$ scattering state or tetraquark molecular state with the  $J^{PC}=1^{+-}$, while the neutral
   current $J_{\mu\nu}(x)$ couples potentially to the $D_s^*\bar{D}_{s1}-D_{s1}\bar{D}_s^*$ meson-meson scattering states or tetraquark molecular states with the $J^{PC}=1^{++}$ and $1^{-+}$. Thereafter, we will denote the charged $D\bar{D}^*+D^*\bar{D}$  tetraquark molecular state with the  $J^{PC}=1^{+-}$ as the $Z_c$, and denote the neutral  $D_s^*\bar{D}_{s1}-D_{s1}\bar{D}_s^*$ tetraquark molecular states with the $J^{PC}=1^{++}$ and $1^{-+}$  as the $X_c^+$ and $X^-_c$, respectively, where the superscripts $\pm$ on the $X_c^\pm$ denote the positive-parity and negative-parity, respectively.

In the following, we write down the possible current-hadron  couplings explicitly,
\begin{eqnarray}\label{JA-MM-Z-MM}
\langle0|J_\mu(0)|D(q)\bar{D}^*(p-q)\rangle&=&\frac{1}{\sqrt{2}}\frac{f_D m_D^2}{m_c}f_{D^*}m_{D^*}\,\varepsilon_\mu(p-q)\, ,\nonumber\\
\langle0|J_\mu(0)|D^*(q)\bar{D}(p-q)\rangle&=&\frac{1}{\sqrt{2}}\frac{f_D m_D^2}{m_c}f_{D^*}m_{D^*}\,\varepsilon_\mu(q)\, ,\nonumber\\
\langle0|J_\mu(0)|D(q)\bar{D}_0(p-q)\rangle&=&\frac{1}{\sqrt{2}}\frac{f_D m_D^2}{m_c}f_{D_0}\,(p-q)_\mu\, ,\nonumber\\
\langle0|J_\mu(0)|D_0(q)\bar{D}(p-q)\rangle&=&\frac{1}{\sqrt{2}}\frac{f_D m_D^2}{m_c}f_{D_0}\,q_\mu\, ,
\end{eqnarray}

\begin{eqnarray}\label{JA-MM-Z}
 \langle 0|J_{\mu}(0)|Z_c(p)\rangle &=& \lambda_{Z} \,  \varepsilon_{\mu}(p)\, ,
\end{eqnarray}

\begin{eqnarray}\label{JT-MM-Z-MM}
\langle0|J_{\mu\nu}(0)|D^*_s(q)\bar{D}_{s1}(p-q)\rangle&=&\frac{1}{\sqrt{2}}f_{D^*_s}m_{D^*_s} f_{D_{s1}}m_{D_{s_1}}\,\varepsilon_{\mu}(q)\varepsilon_\nu(p-q)\, ,\nonumber\\
\langle0|J_{\mu\nu}(0)|D_{s1}(q)\bar{D}^*_{s}(p-q)\rangle&=&-\frac{1}{\sqrt{2}}f_{D_{s1}}m_{D_{s_1}}f_{D^*_s}m_{D^*_s}\,\varepsilon_{\nu}(q)\varepsilon_{\mu}(p-q) \, ,\nonumber\\
\langle0|J_{\mu\nu}(0)|D^*_s(q)\bar{D}_{s}(p-q)\rangle&=&\frac{i}{\sqrt{2}}f_{D^*_s}m_{D^*_s} f_{D_{s}}\,\varepsilon_{\mu}(q)(p-q)_\nu\, ,\nonumber\\
\langle0|J_{\mu\nu}(0)|D_{s}(q)\bar{D}^*_{s}(p-q)\rangle&=&-\frac{i}{\sqrt{2}}f_{D_{s}}f_{D^*_s}m_{D^*_s}\,q_{\nu}\varepsilon_{\mu}(p-q) \, ,\nonumber\\
\langle0|J_{\mu\nu}(0)|D_{s0}(q)\bar{D}_{s}(p-q)\rangle&=&\frac{i}{\sqrt{2}}f_{D_{s0}} f_{D_{s}}\,q_{\mu}(p-q)_\nu\, ,\nonumber\\
\langle0|J_{\mu\nu}(0)|D_{s}(q)\bar{D}_{s0}(p-q)\rangle&=&-\frac{i}{\sqrt{2}}f_{D_{s}}f_{D_{s0}}\,q_{\nu}(p-q)_{\mu} \, ,
\end{eqnarray}

\begin{eqnarray}\label{JT-MM-Z}
  \langle 0|J_{\mu\nu}(0)|X_c^-(p)\rangle &=& \frac{\lambda_{X^-}}{M_{X^-}} \, \varepsilon_{\mu\nu\alpha\beta} \, \varepsilon^{\alpha}(p)p^{\beta}\, , \nonumber\\
 \langle 0|J_{\mu\nu}(0)|X_c^+(p)\rangle &=&\frac{\lambda_{X^+}}{M_{X^+}} \left[\varepsilon_{\mu}(p)p_{\nu}-\varepsilon_{\nu}(p)p_{\mu} \right]\, ,
\end{eqnarray}
the  $\varepsilon_\mu$ are the polarization vectors of the vector and axialvector mesons or tetraquark molecular states,
the $f_{D}$, $f_{D_s}$, $f_{D^*}$, $f_{D_s^*}$, $f_{D_0}$, $f_{D_{s0}}$ and $f_{D_{s1}}$ are the decay constants of the traditional or normal heavy mesons, the $\lambda_Z$ and $\lambda_{X^\pm}$ are the pole residues of the tetraquark molecular states.
The charged $D\bar{D}^*+D^*\bar{D}$  tetraquark molecular state $Z_c$ with the  $J^{PC}=1^{+-}$ and the neutral $D_s^*\bar{D}_{s1}-D_{s1}\bar{D}_s^*$  tetraquark molecular state $X_c^-$ with the $J^{PC}=1^{-+}$ differ from the traditional mesons significantly, and  are good subjects to study the exotic states.

Now we take a short digression to give some explanations  for the definitions of the current-hadron  couplings in Eq.\eqref{JA-MM-Z-MM} and Eq.\eqref{JT-MM-Z-MM}.
Firstly, let us write down the standard  definitions for the decay constants of the  traditional or normal heavy mesons,
\begin{eqnarray}\label{decay-constant}
\langle 0|\bar{q}(0)i\gamma_5c(0)|D(q)\rangle&=&\frac{f_D m_D^2}{m_c}\, , \nonumber \\
\langle 0|\bar{q}(0)\gamma_{\mu}c(0)|D^*(q)\rangle&=&f_{D^*} m_{D^*}\varepsilon_{\mu}(q)\, , \nonumber \\
\langle 0|\bar{s}(0)\gamma_{\mu}c(0)|D_s^*(q)\rangle&=&f_{D_s^*} m_{D_s^*}\varepsilon_{\mu}(q)\, , \nonumber \\
\langle 0|\bar{q}(0)\gamma_{\mu}c(0)|D_0(q)\rangle&=&f_{D_0} q_{\mu}\, , \nonumber \\
\langle 0|\bar{s}(0)\gamma_{\mu}c(0)|D_{s0}(q)\rangle&=&f_{D_{s0}} q_{\mu}\, ,\nonumber \\
\langle 0|\bar{s}(0)\gamma_{\mu}\gamma_5c(0)|D_{s1}(q)\rangle&=&f_{D_{s1}} m_{D_{s1}}\varepsilon_{\mu}(q)\, , \nonumber \\
\langle 0|\bar{s}(0)\gamma_{\mu}\gamma_5c(0)|D_s(q)\rangle&=&if_{D_s} q_{\mu}\, ,
\end{eqnarray}
based on the properties of the vector currents and axialvector currents and their conservation  features, where $q=u$, $d$. On the other hand, the heavy meson fields have the properties,
 \begin{eqnarray}\label{heavy-meson-field}
\langle 0|D_{(s/0/s0)}(0)|D_{(s/0/s0)}(q)\rangle&=&1\, , \nonumber \\
\langle 0|D^*_{(s)\mu}(0)|D_{(s)}^*(q)\rangle&=&\varepsilon_{\mu}(q)\, , \nonumber \\
\langle 0|D_{s1,\mu}(0)|D_{s1}(q)\rangle&=&\varepsilon_{\mu}(q)\, ,
\end{eqnarray}
which imply that $\bar{q}(x)i\gamma_5c(x)=\frac{f_D m_D^2}{m_c}D(x)+\cdots$, etc, at the hadron degrees of freedom.
From Eq.\eqref{JA-MM-Z} and Eqs.\eqref{JT-MM-Z}-\eqref{heavy-meson-field}, we can express  the four-quark currents $J_\mu(x)$ and $J_{\mu\nu}(x)$ in terms of the heavy meson fields (in other words,  the Eq.\eqref{JA-MM-Z} and Eqs.\eqref{JT-MM-Z}-\eqref{heavy-meson-field} imply that),
\begin{eqnarray}
J_\mu(x)&=&\frac{1}{\sqrt{2}}\frac{f_D m_D^2}{m_c}f_{D^*}m_{D^*}\left[D^0(x)D^{*-}_\mu(x)+D^{*0}_\mu(x)D^{-}(x)\right]\nonumber\\
&&+\frac{1}{\sqrt{2}}\frac{f_D m_D^2}{m_c}f_{D_0}\left[D^0(x)i\partial_{\mu}D_{0}^-(x)+i\partial_{\mu}D^{0}_{0}(x)D^{-}(x)\right]+\lambda_{Z}Z_{c,\mu}(x)+\cdots\, ,
\end{eqnarray}

\begin{eqnarray}
J_{\mu\nu}(x)&=&\frac{1}{\sqrt{2}}f_{D_s^*}m_{D_s^*}f_{D_{s1}}m_{D_{s1}}\left[D_{s,\mu}^{*+}(x)D^{-}_{s1,\nu}(x) -D_{s1,\nu}^{+}(x)D^{*-}_{s,\mu}(x)\right]\nonumber\\
&&-\frac{1}{\sqrt{2}}f_{D_s^*}m_{D_s^*}f_{D_{s}}\left[D_{s,\mu}^{*+}(x)\partial_{\nu}D_s^{-}(x) -\partial_{\nu}D_{s}^{+}(x)D^{*-}_{s,\mu}(x)\right]\nonumber\\
&&+\frac{1}{\sqrt{2}}f_{D_{s0}}f_{D_{s1}}m_{D_{s1}}\left[i\partial_{\mu} D_{s0}^{+}(x)D^{-}_{s1,\nu}(x) -D_{s1,\nu}^{+}(x)i\partial_{\mu}D^{-}_{s0}(x)\right]\nonumber\\
&&-\frac{1}{\sqrt{2}}f_{D_{s0}}f_{D_{s}}\left[i\partial_{\mu} D_{s0}^{+}(x)\partial_{\nu}D^{-}_{s}(x) -\partial_{\nu}D_{s}^{+}(x)i\partial_{\mu}D^{-}_{s0}(x)\right]\nonumber\\
&&-\frac{\lambda_{X^-}}{M_{X^-}}\varepsilon_{\mu\nu\alpha\beta}i\partial^{\alpha}X_c^{-\beta}(x)
-\frac{\lambda_{X^+}}{M_{X^+}}\left[i\partial_{\mu}X^{+}_{c,\nu}(x)-i\partial_{\nu}X^{+}_{c,\mu}(x)\right]+\cdots\, ,
\end{eqnarray}
according to the assumption of  current-hadron duality.  It is straightforward to obtain the current-hadron couplings in Eq.\eqref{JA-MM-Z-MM} and Eq.\eqref{JT-MM-Z-MM}.

At the hadron side, we  insert  a complete set of intermediate hadronic states with
the same quantum numbers as the current operators $J_\mu(x)$ and $J_{\mu\nu}(x)$  into the
correlation functions $\Pi_{\mu\nu}(p)$ and $\Pi_{\mu\nu\alpha\beta}(p)$ to obtain the hadronic representation
\cite{SVZ79,Reinders85}. We isolate the contributions of the meson-meson scattering states and the lowest axialvector and vector tetraquark states according to Eqs.\eqref{JA-MM-Z-MM}-\eqref{JT-MM-Z},  and
 get the  results,
\begin{eqnarray}
\Pi_{\mu\nu}(p)&=&\Pi(p^2)\left(-g_{\mu\nu} +\frac{p_\mu p_\nu}{p^2}\right) +\cdots\, \, , \\
\Pi_{\mu\nu\alpha\beta}(p)&=&\Pi_{-}(p^2)\left(g_{\mu\alpha}g_{\nu\beta} -g_{\mu\beta}g_{\nu\alpha} -g_{\mu\alpha}\frac{p_{\nu}p_{\beta}}{p^2}-g_{\nu\beta}\frac{p_{\mu}p_{\alpha}}{p^2}+g_{\mu\beta}\frac{p_{\nu}p_{\alpha}}{p^2}+g_{\nu\alpha}\frac{p_{\mu}p_{\beta}}{p^2}\right) \nonumber\\
&&+\Pi_{+}(p^2)\left( -g_{\mu\alpha}\frac{p_{\nu}p_{\beta}}{p^2}-g_{\nu\beta}\frac{p_{\mu}p_{\alpha}}{p^2}+g_{\mu\beta}\frac{p_{\nu}p_{\alpha}}{p^2}+g_{\nu\alpha}\frac{p_{\mu}p_{\beta}}{p^2}\right) \, ,
\end{eqnarray}
where
\begin{eqnarray}
\Pi(p^2)&=&\frac{\lambda_Z^2}{M_Z^2-p^2}+\Pi_{TW}(p^2)+\cdots\, , \nonumber\\
\Pi_{-}(p^2)&=&P_{-}^{\mu\nu\alpha\beta}\Pi_{\mu\nu\alpha\beta}(p) =\frac{\lambda_{X^-}^2}{M_{X^-}^2-p^2}+\Pi^-_{TW}(p^2)+\cdots\, , \nonumber\\
\Pi_{+}(p^2)&=&P_{+}^{\mu\nu\alpha\beta}\Pi_{\mu\nu\alpha\beta}(p) =\frac{\lambda_{X^+}^2}{M_{X^+}^2-p^2}+\cdots\, ,
\end{eqnarray}
we project out the components $\Pi_{-}(p^2)$ and $\Pi_{+}(p^2)$ by introducing the operators $P_{-}^{\mu\nu\alpha\beta}$ and $P_{+}^{\mu\nu\alpha\beta}$ respectively,
\begin{eqnarray}
P_{-}^{\mu\nu\alpha\beta}&=&\frac{1}{6}\left( g^{\mu\alpha}-\frac{p^\mu p^\alpha}{p^2}\right)\left( g^{\nu\beta}-\frac{p^\nu p^\beta}{p^2}\right)\, , \nonumber\\
P_{+}^{\mu\nu\alpha\beta}&=&\frac{1}{6}\left( g^{\mu\alpha}-\frac{p^\mu p^\alpha}{p^2}\right)\left( g^{\nu\beta}-\frac{p^\nu p^\beta}{p^2}\right)-\frac{1}{6}g^{\mu\alpha}g^{\nu\beta}\, ,
\end{eqnarray}
\begin{eqnarray}\label{Pi-TW}
\Pi_{TW}(p^2)&=&\frac{\lambda_{DD^*}^2}{16\pi^2}\int_{\Delta_1^2}^{s_0}ds \frac{1}{s-p^2}\frac{\sqrt{\lambda(s,m_{D}^2,m_{D^*}^2)}}{s}\left[1+\frac{\lambda(s,m_{D}^2,m_{D^*}^2)}{12sm_{D^*}^2} \right]\nonumber\\
&&+\frac{\lambda_{DD_0}^2}{16\pi^2}\int_{\Delta_2^2}^{s_0}ds \frac{1}{s-p^2}\frac{\sqrt{\lambda(s,m_{D}^2,m_{D_0}^2)}}{s}\frac{\lambda(s,m_{D}^2,m_{D_0}^2)}{12s}
+\cdots\, ,
\end{eqnarray}

\begin{eqnarray}\label{Pi-TW-N}
\Pi^-_{TW}(p^2)&=&\frac{\lambda_{D^*_sD_{s1}}^2}{16\pi^2}\int_{\Delta_3^2}^{s_0}ds \frac{1}{s-p^2}\frac{\sqrt{\lambda(s,m_{D_s^*}^2,m_{D_{s1}}^2)}}{s}\left[1+\frac{\lambda(s,m_{D^*_s}^2,m_{D_{s1}}^2)}{12sm_{D^*_s}^2}  \frac{\lambda(s,m_{D^*_s}^2,m_{D_{s1}}^2)}{12sm_{D_{s1}}^2}\right] \nonumber\\
&&+\frac{\lambda_{D_s^*D_s}^2}{16\pi^2}\int_{\Delta_4^2}^{s_0}ds \frac{1}{s-p^2}\frac{\sqrt{\lambda(s,m_{D_s^*}^2,m_{D_s}^2)}}{s}\frac{\lambda(s,m_{D_s^*}^2,m_{D_s}^2)}{12s}\nonumber\\
&&+\frac{\lambda_{D_{s0}D_{s1}}^2}{16\pi^2}\int_{\Delta_5^2}^{s_0}ds \frac{1}{s-p^2}\frac{\sqrt{\lambda(s,m_{D_{s0}}^2,m_{D_{s1}}^2)}}{s}\frac{\lambda(s,m_{D_{s0}}^2,m_{D_{s1}}^2)}{12s}
+\cdots\, ,
\end{eqnarray}

\begin{eqnarray}
\lambda_{DD^*}^2&=&\frac{f_{D}^2m_{D}^4f_{D^*}^2m_{D^*}^2}{m_c^2}\, ,\nonumber\\
\lambda_{DD_0}^2&=&\frac{f_{D}^2m_{D}^4f_{D_0}^2}{m_c^2}\, ,\nonumber\\
\lambda_{D_s^*D_{s1}}^2&=&f_{D^*_s}^2m_{D^*_s}^2f_{D_{s1}}^2m_{D_{s1}}^2\, ,\nonumber\\
\lambda_{D_s^*D_{s}}^2&=&f_{D^*_s}^2m_{D^*_s}^2f_{D_{s}}^2\, ,\nonumber\\
\lambda_{D_{s0}D_{s1}}^2&=&f_{D_{s0}}^2f_{D_{s1}}^2m_{D_{s1}}^2\, ,
\end{eqnarray}

\begin{eqnarray}
\Delta_1^2&=&(m_{D}+m_{D^*})^2\, ,\nonumber\\
\Delta_2^2&=&(m_{D}+m_{D_0})^2\, ,\nonumber\\
\Delta_3^2&=&(m_{D^*_s}+m_{D_{s1}})^2\, ,\nonumber\\
\Delta_4^2&=&(m_{D_s^*}+m_{D_s})^2\, ,\nonumber\\
\Delta_5^2&=&(m_{D_{s0}}+m_{D_{s1}})^2\, ,
\end{eqnarray}
$\lambda(a,b,c)=a^2+b^2+c^2-2ab-2bc-2ca$. The components $\Pi_{-}(p^2)$ and $\Pi_{+}(p^2)$ receive contributions from the  $D_s^*\bar{D}_{s1}-D_{s1}\bar{D}_s^*$ meson-meson scattering states or tetraquark molecular states with the $J^{PC}=1^{-+}$ and $1^{++}$, respectively.
The conventional hidden-flavor mesons have the normal quantum numbers, $J^{PC}=0^{-+}$, $0^{++}$, $1^{--}$, $1^{+-}$, $1^{++}$, $2^{--}$, $2^{-+}$, $2^{++}$, $\cdots$.
The component $\Pi_{-}(p^2)$ receives  contributions with the exotic quantum numbers $J^{PC}=1^{-+}$, while  the component $\Pi_{+}(p^2)$  receives contributions with the normal quantum numbers $J^{PC}=1^{++}$. In this article,   we study the tetraquark molecular states (in other words, the exotic states), it is better to choose the component $\Pi_{-}(p^2)$ with the exotic quantum numbers $J^{PC}=1^{-+}$, so we discard the  component $\Pi_{+}(p^2)$ with the normal quantum numbers $J^{PC}=1^{++}$. Thereafter, we will neglect  the superscript $-$ in the $X_c^-$ for simplicity.

\begin{figure}
 \centering
  \includegraphics[totalheight=5cm,width=7cm]{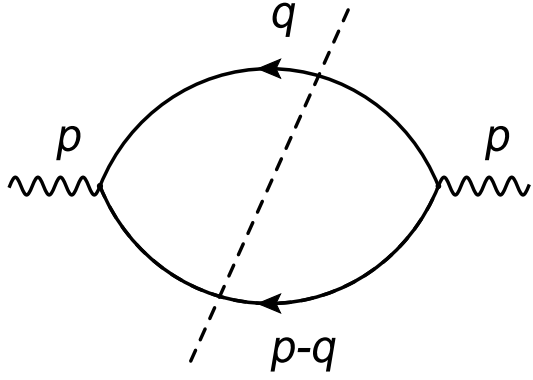}
 \caption{ The  Feynman diagrams   for the two-meson intermediate states, where the  dashed line represents  the cut. }\label{two-meson-cut}
\end{figure}

Now we give some explanations for the components $\Pi_{TW}(p^2)$ and $\Pi^-_{TW}(p^2)$ in Eqs.\eqref{Pi-TW}-\eqref{Pi-TW-N}. We draw up the Feynman diagrams for the two-meson scattering  state contributions in the correlation functions $\Pi_{\mu\nu}(p)$ and $\Pi_{\mu\nu\alpha\beta}(p)$, see Fig.\ref{two-meson-cut}, and  resort to the Cutkosky's rule to calculate the imaginary parts ${\rm Im}\,\Pi_{TW}(p^2)$ and ${\rm Im}\,\Pi^-_{TW}(p^2)$ with the simple replacements of the
two heavy-meson lines,
\begin{eqnarray}
\frac{1}{q^2-m_A^2+i\epsilon} &\to & -2\pi i\,\delta \left( q^2-m_A^2\right)\, , \nonumber\\
\frac{1}{(p-q)^2-m_B^2+i\epsilon} &\to & -2\pi i\,\delta \left( (p-q)^2-m_B^2\right)\, ,
\end{eqnarray}
where the $m_A$ and $m_B$ denote the masses of the two heavy mesons, respectively. Then it is straight forward to carry out the integral over the four-vector $q_\alpha$, and   obtain the two-meson scattering state contributions $\Pi_{TW}(p^2)$ and $\Pi^-_{TW}(p^2)$ through dispersion relation.

 In this article, we carry out the
operator product expansion to the vacuum condensates  up to dimension-10, and take into account the vacuum condensates which are vacuum expectations of the quark-gluon operators  of the order $\mathcal{O}(\alpha_s^k)$ with $k\leq1$. In calculations, we assume  vacuum saturation for the  higher dimensional  vacuum condensates.
For the current $J_\mu(x)$, we take into account the vacuum  condensates
$\langle \bar{q}q\rangle$,
 $\langle \frac{\alpha_s}{\pi}GG\rangle$,
 $\langle \bar{q} g_s \sigma Gq\rangle$,
 $\langle \bar{q}q\rangle^2$, $g_s^2\langle \bar{q}q\rangle^2$,
 $\langle \bar{q}q\rangle\langle \frac{\alpha_s}{\pi}GG\rangle$,
 $\langle \bar{q}q\rangle\langle \bar{q} g_s \sigma Gq\rangle$,
$\langle \bar{q}q\rangle^2\langle \frac{\alpha_s}{\pi}GG\rangle$,
$\langle \bar{q} g_s \sigma Gq\rangle^2$.  The four-quark condensate $g_s^2\langle \bar{q}q\rangle^2$ comes from the terms
$\langle \bar{q}\gamma_\mu t^a q g_s D_\eta G^a_{\lambda\tau}\rangle$, $\langle\bar{q}_jD^{\dagger}_{\mu}D^{\dagger}_{\nu}D^{\dagger}_{\alpha}q_i\rangle$  and
$\langle\bar{q}_jD_{\mu}D_{\nu}D_{\alpha}q_i\rangle$, rather than comes from the perturbative corrections of the $\langle \bar{q}q\rangle^2$. The four-quark condensate $g_s^2\langle \bar{q}q\rangle^2$ plays  an important role in choosing the input parameters due to the relation $g_s^2=4\pi \alpha_s(\mu)$, which introduces explicit
energy scale dependence, on the other hand, it plays a minor important role in numerical calculations.
For the current $J_{\mu\nu}(x)$, we take into account the vacuum  condensates
$\langle \bar{s}s\rangle$,
 $\langle \frac{\alpha_s}{\pi}GG\rangle$,
 $\langle \bar{s} g_s \sigma Gs\rangle$,
 $\langle \bar{s}s\rangle^2$,
 $\langle \bar{s}s\rangle\langle \frac{\alpha_s}{\pi}GG\rangle$,
 $\langle \bar{s}s\rangle\langle \bar{s} g_s \sigma Gs\rangle$,
$\langle \bar{s}s\rangle^2\langle \frac{\alpha_s}{\pi}GG\rangle$,
$\langle \bar{s} g_s \sigma Gs\rangle^2$, and neglect the condensate $g_s^2\langle\bar{s}s\rangle^2$. After carrying out the operator product expansion, we obtain the analytical expressions of the correlation functions $\Pi(p^2)$ and $\Pi_{-}(p^2)$ at the quark-gluon level,
\begin{eqnarray}
\Pi(p^2)&=&\int_{4m_c^2}^{s_0} \frac{\rho_{Z,QCD}(s)}{s-p^2}+\cdots\, , \nonumber\\
\Pi_{-}(p^2)&=&\int_{4m_c^2}^{s_0} \frac{\rho_{X,QCD}(s)}{s-p^2}+\cdots\, ,
\end{eqnarray}
where the $\rho_{Z,QCD}(s)$ and $\rho_{X,QCD}(s)$ are the QCD spectral densities,
\begin{eqnarray}
\rho_{Z,QCD}(s)&=&\frac{1}{\pi}{\rm Im}\Pi(s+i\epsilon)\mid_{\epsilon \to 0}\, , \nonumber \\
\rho_{X,QCD}(s)&=&\frac{1}{\pi}{\rm Im}\Pi_{-}(s+i\epsilon)\mid_{\epsilon \to 0} \, .
\end{eqnarray}

According to the assertion   of Lucha, Melikhov and Sazdjian \cite{Chu-Sheng-PRD}, all the contributions of the order $\mathcal{O}(\alpha_s^k)$ with $k\leq 1$ are exactly canceled out by the meson-meson scattering states, we can set
\begin{eqnarray}
\Pi(p^2)&=&\Pi_{TW}(p^2)+\cdots\, , \nonumber\\
\Pi_{-}(p^2)&=&\Pi^-_{TW}(p^2)+\cdots\, ,
\end{eqnarray}
at the hadron side, as we carry out the operator product expansion by taking into account only the contributions of the order $\mathcal{O}(\alpha_s^k)$ with $k\leq 1$.
 Now let us   take the
quark-hadron duality below the continuum threshold $s_0$ and saturate the hadron side of the correlation functions with the meson-meson scattering states, then  perform Borel transform  with respect to
the variable $P^2=-p^2$ to obtain  the  QCD sum rules:
\begin{eqnarray}\label{TW-A-QCDSR}
\Pi_{TW}(T^2)&=&\frac{\lambda_{DD^*}^2}{16\pi^2}\int_{\Delta_1^2}^{s_0}ds \frac{\sqrt{\lambda(s,m_{D}^2,m_{D^*}^2)}}{s}\left[1+\frac{\lambda(s,m_{D}^2,m_{D^*}^2)}{12sm_{D^*}^2} \right]\exp\left(-\frac{s}{T^2}\right)\nonumber\\
&&+\frac{\lambda_{DD_0}^2}{16\pi^2}\int_{\Delta_2^2}^{s_0}ds \frac{\sqrt{\lambda(s,m_{D}^2,m_{D_0}^2)}}{s}\frac{\lambda(s,m_{D}^2,m_{D_0}^2)}{12s}\exp\left(-\frac{s}{T^2}\right)\nonumber\\
&=&\kappa\int_{4m_c^2}^{s_0}ds\,\rho_{Z,QCD}(s)\exp\left(-\frac{s}{T^2}\right)\, ,
\end{eqnarray}
\begin{eqnarray}\label{TW-Negative-V-QCDSR}
\Pi^-_{TW}(T^2)&=&\frac{\lambda_{D^*_sD_{s1}}^2}{16\pi^2}\int_{\Delta_3^2}^{s_0}ds \frac{\sqrt{\lambda(s,m_{D_s^*}^2,m_{D_{s1}}^2)}}{s}\left[1+\frac{\lambda(s,m_{D^*_s}^2,m_{D_{s1}}^2)}{12sm_{D^*_s}^2}  \frac{\lambda(s,m_{D^*_s}^2,m_{D_{s1}}^2)}{12sm_{D_{s1}}^2}\right]\exp\left(-\frac{s}{T^2}\right) \nonumber\\
&&+\frac{\lambda_{D_s^*D_s}^2}{16\pi^2}\int_{\Delta_4^2}^{s_0}ds \frac{\sqrt{\lambda(s,m_{D_s^*}^2,m_{D_s}^2)}}{s}\frac{\lambda(s,m_{D_s^*}^2,m_{D_s}^2)}{12s}\exp\left(-\frac{s}{T^2}\right)\nonumber\\
&&+\frac{\lambda_{D_{s0}D_{s1}}^2}{16\pi^2}\int_{\Delta_5^2}^{s_0}ds \frac{\sqrt{\lambda(s,m_{D_{s0}}^2,m_{D_{s1}}^2)}}{s}\frac{\lambda(s,m_{D_{s0}}^2,m_{D_{s1}}^2)}{12s}\exp\left(-\frac{s}{T^2}\right)\nonumber\\
&=&\kappa\int_{4m_c^2}^{s_0}ds\,\rho_{X,QCD}(s)\exp\left(-\frac{s}{T^2}\right)\, ,
\end{eqnarray}
 the   explicit expressions of the QCD spectral densities $\rho_{Z,QCD}(s)$ and $\rho_{X,QCD}(s)$ are given in the Appendix, where we have rewritten the
terms of the forms $\frac{d}{ds}\delta(s-\overline{m}_c^2)$, $\frac{d^2}{ds^2}\delta(s-\overline{m}_c^2)$, $\cdots$, $\frac{d}{ds}\delta(s-\widetilde{m}_c^2)$, $\frac{d^2}{ds^2}\delta(s-\widetilde{m}_c^2)$, $\cdots$ in more concise forms.
We saturate the QCD side of the correlation functions with the two-meson scattering states at the hadron side "by hand" according to the assertion of  Lucha, Melikhov and Sazdjian \cite{Chu-Sheng-PRD}. In Sect.2, we present detailed discussions  to approve that the assertion is questionable, we have to introduce some parameters to evaluate the assertion in practical calculations.
In Eqs.\eqref{TW-A-QCDSR}-\eqref{TW-Negative-V-QCDSR},  we introduce the parameter $\kappa$ to measure the deviations from $1$, if $\kappa\approx1$, we can get the conclusion tentatively that
the meson-meson scattering states can  saturate the QCD sum rules.
Then we  differentiate   Eqs.\eqref{TW-A-QCDSR}-\eqref{TW-Negative-V-QCDSR} with respect to  $\frac{1}{T^2}$,   and obtain two additional  QCD sum rules,
 \begin{eqnarray}\label{TW-A-QCDSR-Dr}
-\frac{d\Pi_{TW}(T^2)}{d(1/T^2)}&=&-\kappa\frac{d}{d(1/T^2)}\int_{4m_c^2}^{s_0}ds\,\rho_{Z,QCD}(s)\exp\left(-\frac{s}{T^2}\right)\, ,
\end{eqnarray}
\begin{eqnarray}\label{TW-Negative-V-QCDSR-Dr}
-\frac{d\Pi^-_{TW}(T^2)}{d(1/T^2)}&=&-\kappa\frac{d}{d(1/T^2)}\int_{4m_c^2}^{s_0}ds\,\rho_{X,QCD}(s)\exp\left(-\frac{s}{T^2}\right)\, .
\end{eqnarray}
Thereafter, we will denote the QCD sum rules in Eqs.\eqref{TW-A-QCDSR-Dr}-\eqref{TW-Negative-V-QCDSR-Dr} as the QCDSR I, and
the QCD sum rules in Eqs.\eqref{TW-A-QCDSR}-\eqref{TW-Negative-V-QCDSR} as the QCDSR II.

On the other hand, if the meson-meson scattering states cannot saturate the QCD sum rules, we have to introduce the tetraquark molecular states to saturate the QCD sum rules,
\begin{eqnarray}\label{TetraQ-A-QCDSR}
\lambda_Z^2\exp\left(-\frac{M_Z^2}{T^2}\right)&=&\int_{4m_c^2}^{s_0}ds\,\rho_{Z,QCD}(s)\exp\left(-\frac{s}{T^2}\right)\, ,
\end{eqnarray}
\begin{eqnarray}\label{TetraQ-Negative-V-QCDSR}
\lambda_X^2\exp\left(-\frac{M_X^2}{T^2}\right)&=&\int_{4m_c^2}^{s_0}ds\,\rho_{X,QCD}(s)\exp\left(-\frac{s}{T^2}\right)\, .
\end{eqnarray}

 We differentiate   Eqs.\eqref{TetraQ-A-QCDSR}-\eqref{TetraQ-Negative-V-QCDSR} with respect to  $\frac{1}{T^2}$,   and obtain two QCD sum rules for the masses of the  tetraquark molecular states,
\begin{eqnarray}\label{TetraQ-A-QCDSR-Dr}
M_Z^2&=&\frac{-\frac{d}{d(1/T^2)}\int_{4m_c^2}^{s_0}ds\,\rho_{Z,QCD}(s)\exp\left(-\frac{s}{T^2}\right)}{\int_{4m_c^2}^{s_0}ds\,\rho_{Z,QCD}(s)\exp\left(-\frac{s}{T^2}\right)}\, ,
\end{eqnarray}
\begin{eqnarray}\label{TetraQ-Negative-V-QCDSR-Dr}
M_X^2&=&\frac{-\frac{d}{d(1/T^2)}\int_{4m_c^2}^{s_0}ds\,\rho_{X,QCD}(s)\exp\left(-\frac{s}{T^2}\right)}{\int_{4m_c^2}^{s_0}ds\,\rho_{X,QCD}(s)\exp\left(-\frac{s}{T^2}\right)}\, .
\end{eqnarray}

\section{Numerical results and discussions}
At the QCD side, we choose  the standard values of the vacuum condensates $\langle
\bar{q}q \rangle=-(0.24\pm 0.01\, \rm{GeV})^3$,  $\langle\bar{s}s \rangle=(0.8\pm0.1)\langle\bar{q}q \rangle$,
$\langle\bar{q}g_s\sigma G q \rangle=m_0^2\langle \bar{q}q \rangle$,
 $\langle\bar{s}g_s\sigma G s \rangle=m_0^2\langle \bar{s}s \rangle$,
$m_0^2=(0.8 \pm 0.1)\,\rm{GeV}^2$,   $\langle \frac{\alpha_s
GG}{\pi}\rangle=(0.33\,\rm{GeV})^4 $    at the energy scale  $\mu=1\, \rm{GeV}$
\cite{SVZ79,Reinders85,ColangeloReview}, and choose the $\overline{MS}$ masses  $m_{c}(m_c)=(1.275\pm0.025)\,\rm{GeV}$  and $m_s(\mu=2\,\rm{GeV})=(0.095\pm0.005)\,\rm{GeV}$
 from the Particle Data Group \cite{PDG}, and set $m_u=m_d=0$.
Moreover, we take into account the energy-scale dependence of  the input  parameters,
\begin{eqnarray}
\langle\bar{q}q \rangle(\mu)&=&\langle\bar{q}q \rangle({\rm 1 GeV})\left[\frac{\alpha_{s}({\rm 1 GeV})}{\alpha_{s}(\mu)}\right]^{\frac{12}{25}}\, , \nonumber\\
 \langle\bar{s}s \rangle(\mu)&=&\langle\bar{s}s \rangle({\rm 1GeV})\left[\frac{\alpha_{s}({\rm 1GeV})}{\alpha_{s}(\mu)}\right]^{\frac{12}{25}}\, , \nonumber\\
 \langle\bar{q}g_s \sigma Gq \rangle(\mu)&=&\langle\bar{q}g_s \sigma Gq \rangle({\rm 1 GeV})\left[\frac{\alpha_{s}({\rm 1 GeV})}{\alpha_{s}(\mu)}\right]^{\frac{2}{25}}\, , \nonumber\\
 \langle\bar{s}g_s \sigma G s \rangle(\mu)&=&\langle\bar{s}g_s \sigma Gs \rangle({\rm 1GeV})\left[\frac{\alpha_{s}({\rm 1GeV})}{\alpha_{s}(\mu)}\right]^{\frac{2}{25}}\, , \nonumber\\
m_c(\mu)&=&m_c(m_c)\left[\frac{\alpha_{s}(\mu)}{\alpha_{s}(m_c)}\right]^{\frac{12}{25}} \, ,\nonumber\\
m_s(\mu)&=&m_s({\rm 2GeV} )\left[\frac{\alpha_{s}(\mu)}{\alpha_{s}({\rm 2GeV})}\right]^{\frac{12}{25}} \, ,\nonumber\\
\alpha_s(\mu)&=&\frac{1}{b_0t}\left[1-\frac{b_1}{b_0^2}\frac{\log t}{t} +\frac{b_1^2(\log^2{t}-\log{t}-1)+b_0b_2}{b_0^4t^2}\right]\, ,
\end{eqnarray}
   where $t=\log \frac{\mu^2}{\Lambda^2}$, $b_0=\frac{33-2n_f}{12\pi}$, $b_1=\frac{153-19n_f}{24\pi^2}$,
   $b_2=\frac{2857-\frac{5033}{9}n_f+\frac{325}{27}n_f^2}{128\pi^3}$,
   $\Lambda=210\,\rm{MeV}$, $292\,\rm{MeV}$  and  $332\,\rm{MeV}$ for the flavors
   $n_f=5$, $4$ and $3$, respectively  \cite{PDG,Narison-mix}, and evolve all the input  parameters to the ideal   energy scales    $\mu$  with $n_f=4$ to extract the
    tetraquark molecular masses or the parameters $\kappa$. The QCD spectral densities $\rho_{Z,QCD}(s)$ and $\rho_{X,QCD}(s)$, and the thresholds $4m_c^2$  depend on
     the energy scales $\mu$,  the values of the parameters $\kappa$, masses $M_{Z/X}$ and pole residues $\lambda_{Z/X}$ extracted from  the QCD sum rules in Eqs.\eqref{TW-A-QCDSR}-\eqref{TetraQ-Negative-V-QCDSR-Dr} vary with the energy scales $\mu$, we should resort to some methods to choose the ideal energy scales (or pertinent energy scales) $\mu$ to extract  those quantities in a consistent way.

At the hadron side, we take the hadronic parameters  as
$m_{D}=1.8672\,\rm{GeV}$, $m_{D_s}=1.9690\,\rm{GeV}$,
$m_{D^*}=2.0086\,\rm{GeV}$, $m_{D_s^*}=2.1122\,\rm{GeV}$,
$m_{D_0}=2.3245\,\rm{GeV}$, $m_{D_{s0}}=2.3180\,\rm{GeV}$, $m_{D_{s1}}=2.5352\,\rm{GeV}$
 from the Particle Data Group \cite{PDG};
 $f_{D}=0.208 \,\rm{GeV}$, $f_{D_s}=0.240 \,\rm{GeV}$,
$f_{D^*}=0.263 \,\rm{GeV}$, $f_{D_s^*}=0.308 \,\rm{GeV}$,
$f_{D_0}=0.373 \,\rm{GeV}$, $f_{D_{s0}}=0.333 \,\rm{GeV}$ \cite{Wang-DecayConst}, $f_{D_{s1}}=0.364 \,\rm{GeV}$ from the QCD sum rules.

The $D\bar{D}^*+D^*\bar{D}$ and $D_s^*\bar{D}_{s1}-D_{s1}\bar{D}_s^*$  thresholds are $m_{D}+m_{D^*}=3.88\,\rm{GeV}$ and $m_{D_s^*}+m_{D_{s1}}=4.65\,\rm{GeV}$, respectively. For the conventional heavy mesons, the mass-gaps between the ground states and the first radial excited states are about $0.4-0.6\,\rm{GeV}$, so the continuum threshold parameters can be chosen as $\sqrt{s_0}=4.40\pm0.10\,\rm{GeV}$ and $5.15\pm0.10\,\rm{GeV}$, respectively.

We  search for the acceptable  Borel parameters $T^2$ to warrant convergence of the operator product expansion and  pole dominance  via trial  and error.
Firstly, let us  define the pole contributions $\rm{PC}$,
\begin{eqnarray}
{\rm PC}&=& \frac{ \int_{4m_c^2}^{s_0} ds\,\rho_{Z/X,QCD}(s)\,\exp\left(-\frac{s}{T^2}\right)}{\int_{4m_c^2}^{\infty} ds \,\rho_{Z/X,QCD}(s)\,\exp\left(-\frac{s}{T^2}\right)}\, ,
\end{eqnarray}
and the contributions of the vacuum condensates $D(n)$,
\begin{eqnarray}
D(n)&=& \frac{  \int_{4m_c^2}^{s_0} ds\,\rho_{Z/X,QCD;n}(s)\,\exp\left(-\frac{s}{T^2}\right)}{\int_{4m_c^2}^{s_0} ds \,\rho_{Z/X,QCD}(s)\,\exp\left(-\frac{s}{T^2}\right)}\, ,
\end{eqnarray}
 where the subscript $n$ in the QCD spectral densities  $\rho_{Z/X,QCD;n}(s)$  represents  the vacuum condensates  of dimension $n$.

\begin{figure}
\centering
\includegraphics[totalheight=7cm,width=9cm]{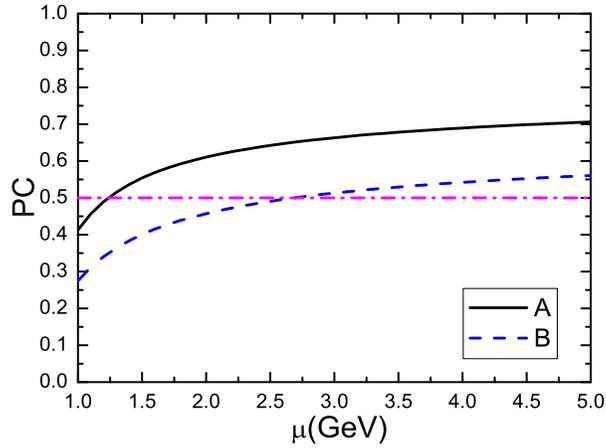}
  \caption{ The pole contributions   with variations of the  energy scales $\mu$, where the $A$ and $B$ correspond to the QCD spectral densities $\rho_{Z,QCD}(s)$ and $\rho_{X,QCD}(s)$, respectively. }\label{PC-mu}
\end{figure}

In Fig.\ref{PC-mu}, we plot the pole contributions with variations of the energy scales of the QCD spectral densities with the parameters  $T_Z^2=2.9\,\rm{GeV}^2$, $\sqrt{s^0_Z}=4.40\,\rm{GeV}$  and $T_X^2=3.9\,\rm{GeV}^2$, $\sqrt{s^0_X}=5.15\,\rm{GeV}$ for the QCD spectral densities $\rho_{Z,QCD}(s)$ and $\rho_{X,QCD}(s)$, respectively.
We choose those typical values because the continuum threshold parameters $s_0$ and Borel parameters $T^2$ have the relation $\frac{s^0_{Z}}{T_Z^2}=\frac{s^0_X}{T_X^2}$, the weight functions $\exp\left(-\frac{s}{T^2}\right)$ have the same values. From Fig.\ref{PC-mu}, we can see that the pole contributions increase monotonically and
 considerably with the increase of the energy scales at the region  $\mu<3.0\,\rm{GeV}$, then the pole contributions increase monotonically but slowly
 with the increase of the energy scales. The pole contributions exceed $50\%$ at the energy scales $\mu=1.3\,\rm{GeV}$ and $2.7\,\rm{GeV}$ for the QCD spectral densities $\rho_{Z,QCD}(s)$ and $\rho_{X,QCD}(s)$, respectively.

\begin{figure}
\centering
\includegraphics[totalheight=7cm,width=9cm]{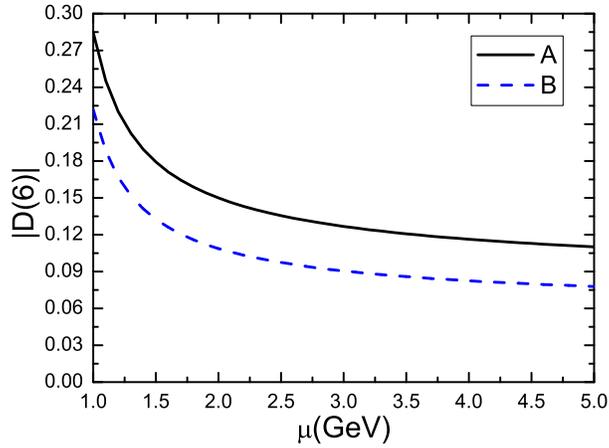}
  \caption{ The absolute values of the $D(6)$   with variations of the  energy scales $\mu$, where the $A$ and $B$ correspond to the QCD spectral densities $\rho_{Z,QCD}(s)$ and $\rho_{X,QCD}(s)$, respectively. }\label{D6-mu}
\end{figure}

In Fig.\ref{D6-mu}, we plot the absolute values of the $D(6)$ with variations of the energy scales $\mu$ of the QCD spectral densities with the parameters  $T_Z^2=2.9\,\rm{GeV}^2$, $\sqrt{s^0_Z}=4.40\,\rm{GeV}$  and $T_X^2=3.9\,\rm{GeV}^2$, $\sqrt{s^0_X}=5.15\,\rm{GeV}$ for the QCD spectral densities $\rho_{Z,QCD}(s)$ and $\rho_{X,QCD}(s)$, respectively. The contributions of the vacuum condensates of dimension $6$ play a very important role in the QCD sum rules for the hidden-charm or hidden-bottom tetraquark (molecular) states.  From Fig.\ref{D6-mu}, we can see that the  contributions $|D(6)|$ decrease monotonically  with the increase of the energy scales. A larger energy scale $\mu$ leads to a larger pole contribution, but a smaller contribution of the vacuum condensate $D(6)$. Too small
contributions of the  vacuum condensates will impair the stability of the QCD sum rules.

\subsection{Meson-meson scattering states alone}
We saturate the hadron side of the QCD sum rules with the meson-meson scattering states alone, and study the QCD sum rues shown in Eqs.\eqref{TW-A-QCDSR}-\eqref{TW-Negative-V-QCDSR-Dr}. In this article, we choose the pole contributions as large as  $(40-60)\%$, the pole dominance criterion is satisfied.
The Borel windows, continuum threshold parameters, energy scales of the QCD spectral densities and pole contributions are shown explicitly in Table \ref{kappa}.
In the Borel windows, the contributions of the higher dimensional  vacuum condensates are $|D(8)|=(3-5)\%$, $|D(10)|\ll1\%$ and
$D(8)=(1-2)\%$, $D(10)\ll1\%$ for the QCD spectral densities $\rho_{Z,QCD}(s)$ and $\rho_{X,QCD}(s)$, respectively. The operator product expansion converges very very good.

\begin{table}
\begin{center}
\begin{tabular}{|c|c|c|c|c|c|c|c|}\hline\hline
$J^{PC}$                     &$T^2 (\rm{GeV}^2)$ &$\sqrt{s_0} (\rm{GeV})$ &$\mu(\rm GeV)$ &pole         & $\kappa_{\rm I}$        & $\kappa_{\rm II}$ \\ \hline

$1^{+-}\,(\bar{u}c\bar{c}d)$ &$2.7-3.1$          &$4.40\pm0.10$           &$1.3$          &$(40-63)\%$  & $1.55\pm0.40$           & $1.37\pm0.40$ \\ \hline

$1^{-+}\,(\bar{s}c\bar{c}s)$ &$3.7-4.1$          &$5.15\pm0.10$           &$2.9$          &$(42-60)\%$  & $0.50\pm0.09$           & $0.46\pm0.09$ \\ \hline
 \hline
\end{tabular}
\end{center}
\caption{ The Borel parameters, continuum threshold parameters, energy scales of the QCD spectral densities, pole contributions and $\kappa$ for the QCDSR I and II, where we show the quark constituents of the meson-meson scattering  states in the brackets. }\label{kappa}
\end{table}

We take into account all uncertainties of the input parameters at the QCD side,
and obtain the values of the $\kappa$ from the QCDSR I and II directly, which are  shown in Table \ref{kappa}. In calculations, we add an uncertainty $\delta \mu=\pm0.1\,\rm{GeV}$ to the energy scales $\mu$. From Table \ref{kappa}, we can see that  the values
 $\kappa_{\rm I}=1.55\pm0.40$   and      $\kappa_{\rm II}=1.37\pm0.40$  overestimate  the contributions of the  $\bar{u}c\bar{c}d$ meson-meson scattering states with the  $J^{PC}=1^{+-}$, while the values $\kappa_{\rm I}=0.50\pm0.09$   and      $\kappa_{\rm II}=0.46\pm0.09$  underestimate the contributions of the  $\bar{s}c\bar{c}s$ meson-meson scattering states with the $J^{PC}=1^{-+}$. In the two cases, the values of the $\kappa$ from  the QCDSR I and II deviate from $1$ significantly, the two-meson scattering sates cannot saturate the QCD sum rules.

\begin{figure}
\centering
\includegraphics[totalheight=6cm,width=7cm]{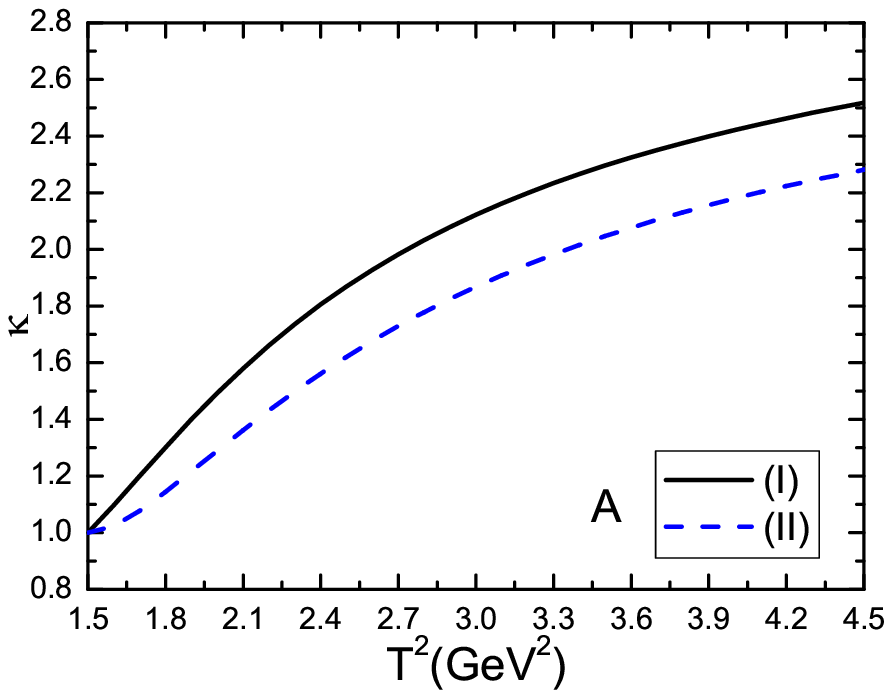}
\includegraphics[totalheight=6cm,width=7cm]{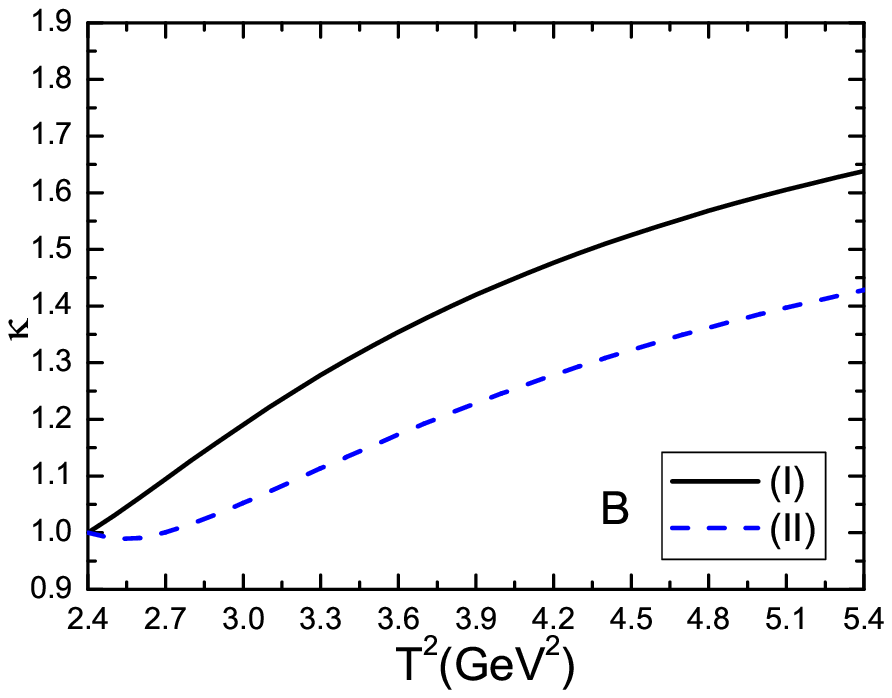}
  \caption{ The $\kappa$  with variations of the  Borel parameters $T^2$, where the $A$ and $B$ correspond to the $\bar{u}c\bar{c}d$ and  $\bar{s}c\bar{c}s$ meson-meson scattering states, respectively,  the (I) and (II) correspond to  QCDSR I and II, respectively, the  $\kappa$ values are normalized to be $1$ for the Borel parameters $T^2=1.5\,\rm{GeV}^2$ and $2.4\,\rm{GeV}^2$, respectively. }\label{kappa-Borel}
\end{figure}

In Fig.\ref{kappa-Borel}, we  plot the values of the $\kappa$    with variations of the  Borel parameters $T^2$ with the continuum threshold parameters
   $\sqrt{s_0}=4.40\,\rm{GeV}$  and  $5.15\,\rm{GeV}$ for the $\bar{u}c\bar{c}d$ and  $\bar{s}c\bar{c}s$ meson-meson scattering states, respectively, where we normalize the  values of the $\kappa$   to be $1$ at the points  $T^2=1.5\,\rm{GeV}^2$ and $2.4\,\rm{GeV}^2$  for the $\bar{u}c\bar{c}d$ and  $\bar{s}c\bar{c}s$ meson-meson scattering states, respectively. In this way, we can see the variation trends of the $\kappa$ with the changes of the Borel parameters more explicitly.
From Fig.\ref{kappa-Borel}, we can see that the values of the $\kappa$ increase monotonically and
 quickly  with the increase of the Borel parameters $T^2$, no platform appears, which indicates  that the QCD sum rules in Eqs.\eqref{TW-A-QCDSR}-\eqref{TW-Negative-V-QCDSR} obtained according to the assertion of  Lucha, Melikhov and Sazdjian are unreasonable. Reasonable QCD sum rules lead to platforms flat enough  or not flat enough, rather than  no evidence of platforms.

Now we can obtain the conclusion tentatively that the meson-meson scattering states cannot saturate the QCD sum rules at the hadron side.

\subsection{Tetraquark molecular states alone}
We saturate the hadron side of the QCD sum rules with the tetraquark molecular states alone, and study the QCD sum rues shown in Eqs.\eqref{TetraQ-A-QCDSR}-\eqref{TetraQ-Negative-V-QCDSR-Dr}.

\begin{table}
\begin{center}
\begin{tabular}{|c|c|c|c|c|c|c|c|}\hline\hline
$J^{PC}$                     &$T^2 (\rm{GeV}^2)$ &$\sqrt{s_0} (\rm{GeV})$&$\mu(\rm GeV)$    &pole         & $M(\rm{GeV})$           & $\lambda(10^{-2}\rm{GeV}^5)$ \\ \hline

$1^{+-}\,(\bar{u}c\bar{c}d)$ &$2.7-3.1$          &$4.40\pm0.10$          &$1.3$             &$(40-63)\%$  & $3.89\pm0.09$           & $1.72\pm0.30$ \\ \hline

$1^{-+}\,(\bar{s}c\bar{c}s)$ &$3.7-4.1$          &$5.15\pm0.10$          &$2.9$             &$(42-60)\%$  & $4.67\pm0.08$           & $6.87\pm0.84$ \\ \hline
 \hline
\end{tabular}
\end{center}
\caption{ The Borel windows, continuum threshold parameters, energy scales of the QCD spectral densities, pole contributions, masses and pole residues of the $\bar{u}c\bar{c}d$ and  $\bar{s}c\bar{c}s$ tetraquark molecular states. }\label{Borel-pole-mass}
\end{table}

\begin{figure}
\centering
\includegraphics[totalheight=7cm,width=9cm]{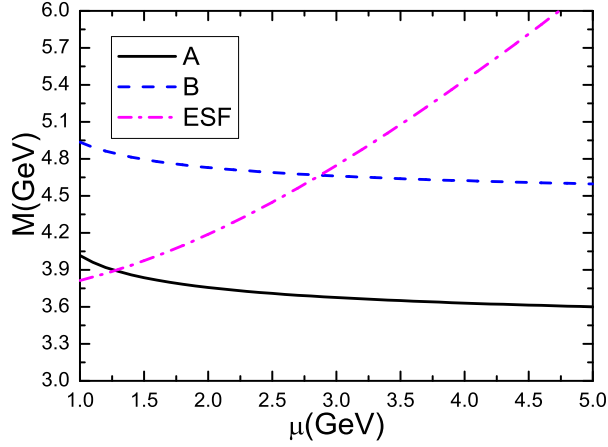}
  \caption{ The masses  with variations of the  energy scales $\mu$, where the $A$ and $B$ correspond to the $\bar{u}c\bar{c}d$ and  $\bar{s}c\bar{c}s$ tetraquark molecular states, respectively, the ESF denotes the energy scale formula $M_{X/Z}=\sqrt{\mu^2+4\times(1.84\,\rm{GeV})^2}$. }\label{mass-mu}
\end{figure}

In Fig.\ref{mass-mu}, we plot the masses with variations of the energy scales of the QCD spectral densities with the parameters  $T_Z^2=2.9\,\rm{GeV}^2$, $\sqrt{s^0_Z}=4.40\,\rm{GeV}$  and $T_X^2=3.9\,\rm{GeV}^2$, $\sqrt{s^0_X}=5.15\,\rm{GeV}$ for the $\bar{u}c\bar{c}d$ and  $\bar{s}c\bar{c}s$ tetraquark molecular states, respectively. From Fig.\ref{mass-mu}, we can see that the values of the masses decrease monotonically and
 slowly  with the increase of the energy scales $\mu$. Now we encounter the problem how to choose the pertinent energy scales of the QCD spectral densities $\rho_{Z,QCD}(s)$ and $\rho_{X,QCD}(s)$.

We describe the heavy tetraquark system $Q\bar{Q}q\bar{q}$ (or the exotic  $X$, $Y$, $Z$ states) by a double-well potential with the two light quarks $q$ and $\bar{q}$ lying in the two potential wells, respectively.
   In the heavy quark limit, the $Q$-quark serves as an  static well potential,
and attracts the light quark $q$ to form a diquark in the color antitriplet channel or attracts the light antiquark $\bar{q}$ to
form a meson in the color singlet channel.
Then the heavy tetraquark (molecular) states  are characterized by the effective heavy quark mass ${\mathbb{M}}_Q$ (or constituent quark mass) and
the virtuality $V=\sqrt{M^2_{X/Y/Z}-(2{\mathbb{M}}_Q)^2}$.
 It is natural to choose  the energy  scales of the QCD spectral densities as,
 \begin{eqnarray}
 \mu^2&=&V^2=M^2_{X/Y/Z}-(2{\mathbb{M}}_Q)^2\, .
 \end{eqnarray}
Analysis of the $J/\psi$ and $\Upsilon$
with the famous Coulomb-plus-linear potential or Cornell potential leads to the constituent quark masses $m_c=1.84\,\rm{GeV}$ and $m_b=5.17\,\rm{GeV}$ \cite{Cornell}.
If we set ${\mathbb{M}}_c=m_c=1.84\,\rm{GeV}$, we can obtain the dash-dotted line $M_{X/Z}=\sqrt{\mu^2+4\times(1.84\,\rm{GeV})^2}$ in Fig.\ref{mass-mu}, which intersects with the lines of the masses of the
$\bar{u}c\bar{c}d$ and  $\bar{s}c\bar{c}s$ tetraquark molecular states at the energy scales about $\mu=1.3\,\rm{GeV}$ and $2.9\,\rm{GeV}$, respectively.
 The values of the cross points are $3.89\,\rm{GeV}$ and $4.67\,\rm{GeV}$, which happen to coincide with the  $D\bar{D}^*+D^*\bar{D}$ and $D_s^*\bar{D}_{s1}-D_{s1}\bar{D}_s^*$  thresholds  $3.88\,\rm{GeV}$ and $4.65\,\rm{GeV}$, respectively. The old value ${\mathbb{M}}_c=1.84\,\rm{GeV}$ and updated value
 ${\mathbb{M}}_c=1.85\,\rm{GeV}$ fitted in the QCD sum rules for the hidden-charm tetraquark molecular states
 are all consistent with the  constituent quark mass $m_c=1.84\,\rm{GeV}$ \cite{WangZG-4-quark-mole,WangZG-CPC-Y4390}.
We can set the value of the effective $c$-quark mass as ${\mathbb{M}}_c=1.84\pm0.01\,\rm{GeV}$. In this article, we use the energy scale formula
$\mu=\sqrt{M_{X/Z}^2-4\times(1.84\,\rm{GeV})^2}$ as the constraints  to choose the best energy scales of the QCD spectral densities.

Again, we choose the pole contributions as large as $(40-60)\%$.
The Borel windows, continuum threshold parameters, energy scales of the QCD spectral densities and pole contributions are shown explicitly in Table \ref{Borel-pole-mass}, just like in Table \ref{kappa}. Again, in the Borel windows,  $|D(8)|=(3-5)\%$, $|D(10)|\ll1\%$ and
$D(8)=(1-2)\%$, $D(10)\ll1\%$ for the  QCD spectral densities $\rho_{Z,QCD}(s)$ and $\rho_{X,QCD}(s)$, respectively.

\begin{figure}
\centering
\includegraphics[totalheight=6cm,width=7cm]{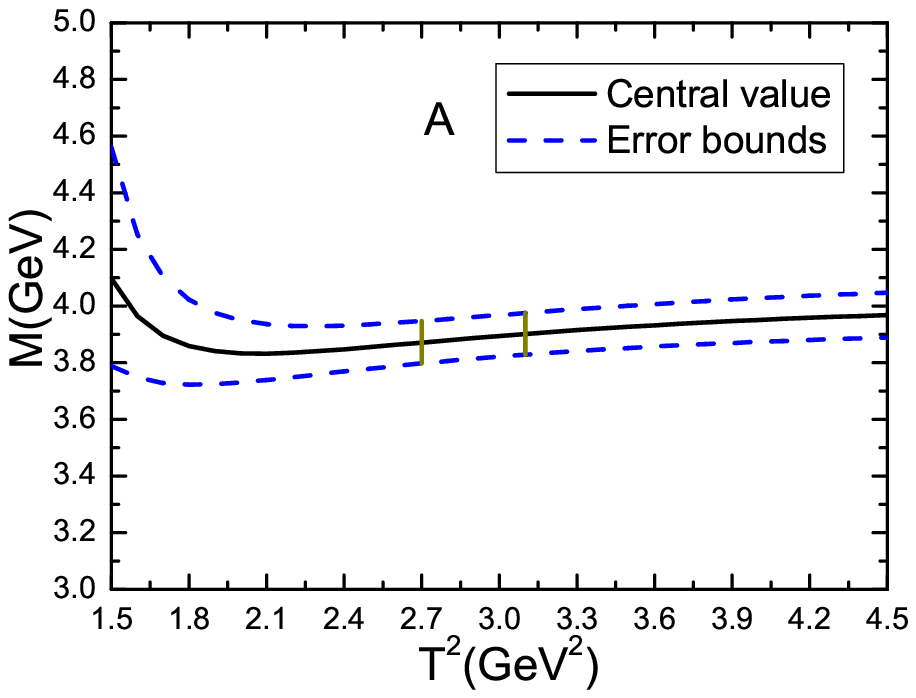}
\includegraphics[totalheight=6cm,width=7cm]{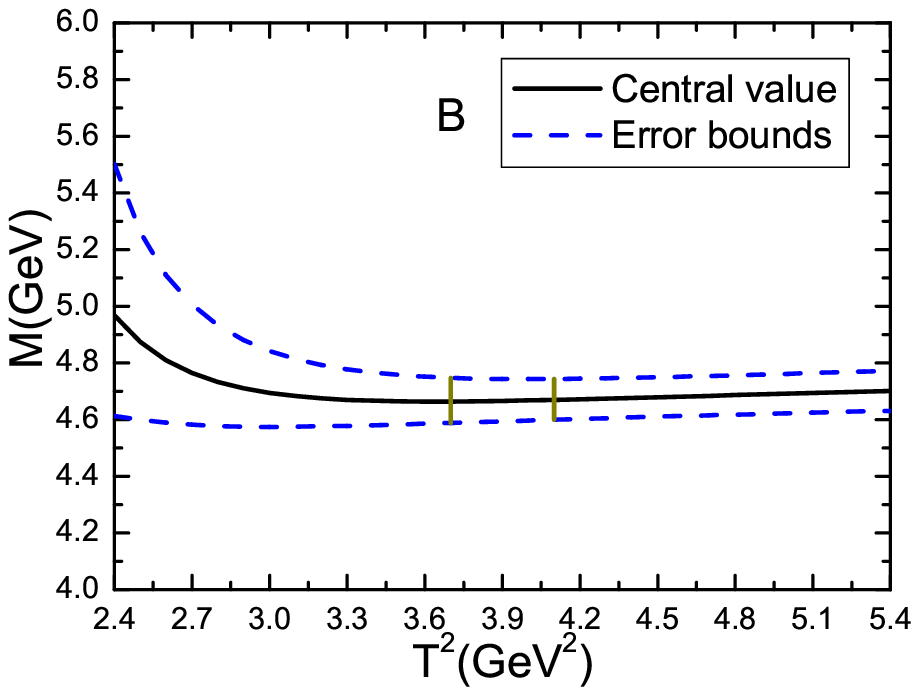}
  \caption{ The masses  with variations of the  Borel parameters $T^2$, where the $A$ and $B$ correspond to the $\bar{u}c\bar{c}d$ and  $\bar{s}c\bar{c}s$ tetraquark molecular states, respectively, the regions between the two vertical lines are the Borel windows. }\label{mass-Borel}
\end{figure}

\begin{figure}
\centering
\includegraphics[totalheight=6cm,width=7cm]{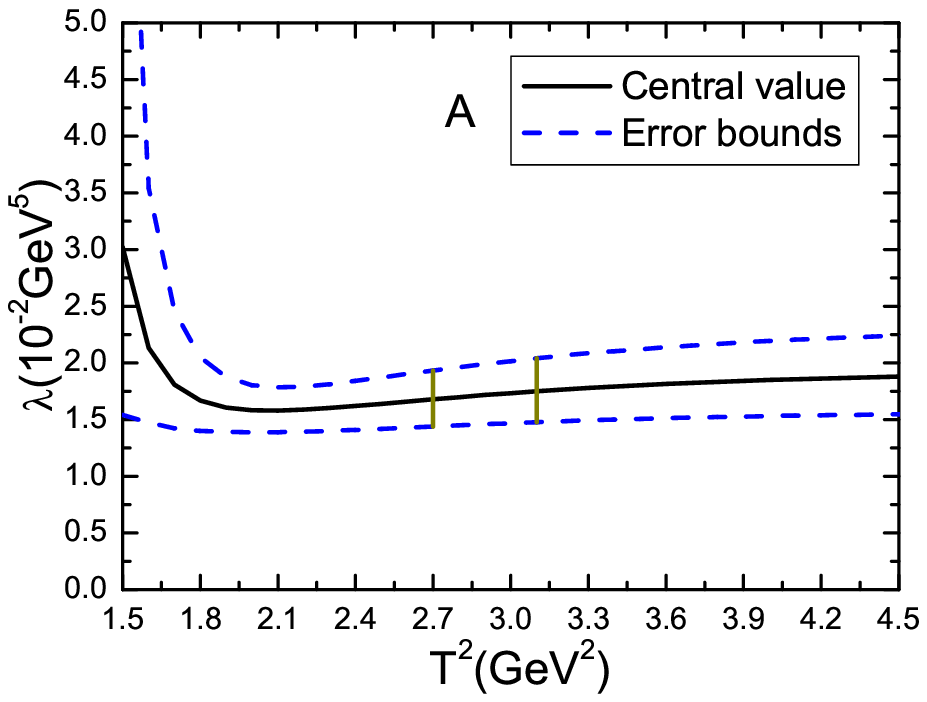}
\includegraphics[totalheight=6cm,width=7cm]{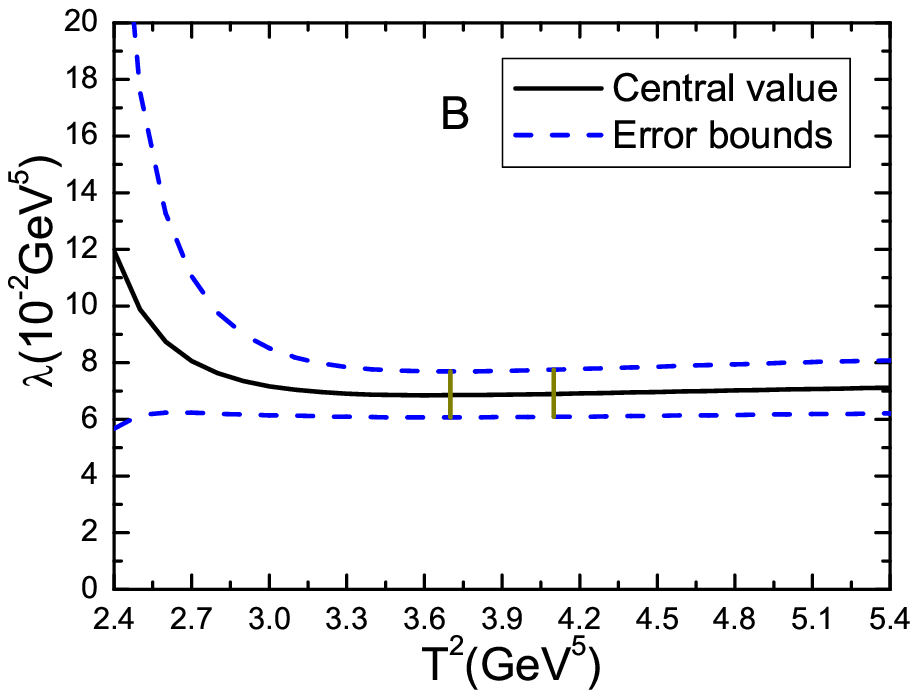}
  \caption{ The pole residues  with variations of the  Borel parameters $T^2$, where the $A$ and $B$ correspond to the $\bar{u}c\bar{c}d$ and  $\bar{s}c\bar{c}s$ tetraquark molecular states, respectively, the regions between the two vertical lines are the Borel windows. }\label{residue-Borel}
\end{figure}

Now let us take into account all uncertainties of the input parameters,
and obtain the values of the masses and pole residues of the tetraquark molecular states, which are  shown in Table \ref{Borel-pole-mass} and Figs.\ref{mass-Borel}-\ref{residue-Borel}. In calculations, we add an uncertainty $\delta \mu=\pm0.1\,\rm{GeV}$ to the energy scales $\mu$ according the
uncertainty in the effective $c$-quark mass ${\mathbb{M}}_c=1.84\pm0.01\,\rm{GeV}$. From Figs.\ref{mass-Borel}-\ref{residue-Borel}, we can see that  there appear Borel platforms in the Borel windows indeed. The tetraquark molecular states alone can satisfy the QCD sum rules.

\subsection{Taking into account the meson-meson scattering states besides the tetraquark molecular states}
From the previous two subsections, we observe that the meson-meson scattering scattering states alone cannot saturate the QCD sum rules, while the tetraquark molecular states alone can saturate the QCD sum rules. However, the quantum field theory  does not forbid the couplings between the four-quark currents and meson-meson scattering states if they have the same quantum numbers. We should take into account both the  tetraquark molecular states and the meson-meson scattering states at the hadron side.

 Now we study the contributions of the intermediate   meson-meson  scattering states  $D
\bar{D}^\ast$, $J/\psi \pi$, $J/\psi \rho$, etc besides the tetraquark molecular state $Z_c$ to the correlation function $\Pi_{\mu\nu}(p)$ as an example,
\begin{eqnarray}\label{Self-Energy}
\Pi_{\mu\nu}(p) &=&-\frac{\widehat{\lambda}_{Z}^{2}}{ p^2-\widehat{M}_{Z}^2-\Sigma_{DD^*}(p^2)-\Sigma_{J/\psi\pi}(p^2)-\Sigma_{J/\psi\rho}(p^2)+\cdots}\widetilde{g}_{\mu\nu}(p)+\cdots \, ,
\end{eqnarray}
where
$\widetilde{g}_{\mu\nu}(p)=-g_{\mu\nu}+\frac{p_{\mu}p_{\nu}}{p^2}$. We choose the bare quantities $\widehat{\lambda}_{Z}$ and $\widehat{M}_{Z}$  to absorb the divergences in the self-energies $\Sigma_{D\bar{D}^*}(p^2)$, $\Sigma_{J/\psi \pi}(p^2)$, $\Sigma_{J/\psi \rho}(p^2)$, etc. The renormalized energies  satisfy  the relation $p^2-M_{Z}^2-\overline{\Sigma}_{DD^*}(p^2)-\overline{\Sigma}_{J/\psi\pi}(p^2)-\overline{\Sigma}_{J/\psi\rho}(p^2)+\cdots=0$, where the overlines above the
self-energies denote that the divergent terms have been subtracted. As the tetraquark molecular state $Z_c$ is unstable, the relation should be modified,
$p^2-M_{Z}^2-{\rm Re}\overline{\Sigma}_{DD^*}(p^2)-{\rm Re}\overline{\Sigma}_{J/\psi\pi}(p^2)-{\rm Re}\overline{\Sigma}_{J/\psi\rho}(p^2)+\cdots=0$, and
$-{\rm Im}\overline{\Sigma}_{DD^*}(p^2)-{\rm Im}\overline{\Sigma}_{J/\psi\pi}(p^2)-{\rm Im}\overline{\Sigma}_{J/\psi\rho}(p^2)+\cdots=\sqrt{p^2}\Gamma(p^2)$. The negative sign in front of the self-energies come from the special definitions in this article, if we redefine the self-energies, $\Sigma(p^2) \to -\Sigma(p^2)$ in Eq.\eqref{Self-Energy}, the negative sign can be removed.
The renormalized self-energies  contribute  a finite imaginary part to modify the dispersion relation,
\begin{eqnarray}
\Pi_{\mu\nu}(p) &=&-\frac{\lambda_{Z}^{2}}{ p^2-M_{Z}^2+i\sqrt{p^2}\Gamma(p^2)}\widetilde{g}_{\mu\nu}(p)+\cdots \, .
 \end{eqnarray}
If we assign the $Z_c(3900)$ to be the $D\bar{D}^*+D^*\bar{D}$  tetraquark molecular state with the $J^{PC}=1^{+-}$ \cite{WangZG-4-quark-mole},  the physical  width $\Gamma_{Z_c(3900)}(M_Z^2)=(28.2 \pm 2.6)\, \rm{MeV}$ from the Particle Data Group \cite{PDG}.

We can take into account the finite width effect by the following simple replacement of the hadronic spectral density,
\begin{eqnarray}
\lambda^2_{Z}\delta \left(s-M^2_{Z} \right) &\to& \lambda^2_{Z}\frac{1}{\pi}\frac{M_{Z}\Gamma_{Z}(s)}{(s-M_{Z}^2)^2+M_{Z}^2\Gamma_{Z}^2(s)}\, ,
\end{eqnarray}
where
\begin{eqnarray}
\Gamma_{Z}(s)&=&\Gamma_{Z} \frac{M_{Z}}{\sqrt{s}}\sqrt{\frac{s-(M_{D}+M_{D^*})^2}{M^2_{Z}-(M_{D}+M_{D^*})^2}} \, .
\end{eqnarray}
Then the hadron  sides of  the QCD sum rules in Eq.\eqref{TetraQ-A-QCDSR} and Eq.\eqref{TetraQ-A-QCDSR-Dr} undergo the following changes,
\begin{eqnarray}
\lambda^2_{Z}\exp \left(-\frac{M^2_{Z}}{T^2} \right) &\to& \lambda^2_{Z}\int_{(m_{D}+m_{D^*})^2}^{s_0}ds\frac{1}{\pi}\frac{M_{Z}\Gamma_{Z}(s)}{(s-M_{Z}^2)^2+M_{Z}^2\Gamma_{Z}^2(s)}\exp \left(-\frac{s}{T^2} \right)\, , \nonumber\\
&=&(0.78\sim0.79)\,\lambda^2_{Z}\exp \left(-\frac{M^2_{Z}}{T^2} \right)\, , \\
\lambda^2_{Z}M^2_{Z}\exp \left(-\frac{M^2_{Z}}{T^2} \right) &\to& \lambda^2_{Z}\int_{(m_{D}+m_{D^*})^2}^{s_0}ds\,s\,\frac{1}{\pi}\frac{M_{Z}\Gamma_{Z}(s)}{(s-M_{Z}^2)^2+M_{Z}^2\Gamma_{Z}^2(s)}\exp \left(-\frac{s}{T^2} \right)\, , \nonumber\\
&=&(0.80\sim0.81)\,\lambda^2_{Z}M^2_{Z}\exp \left(-\frac{M^2_{Z}}{T^2} \right)\, ,
\end{eqnarray}
with the value $\sqrt{s_0}=4.40\,\rm{GeV}$.
We can absorb the numerical factors  $0.78\sim0.79$ and $0.80\sim0.81$ into the pole residue with the simple replacement $\lambda_{Z}\to 0.89\lambda_Z$ safely, the intermediate   meson-loops cannot  affect  the mass $M_{Z}$ significantly, but affect the pole residue remarkably, which are consistent with the fact that we obtain the masses of the tetraquark molecular states from a fraction, see Eqs.\eqref{TetraQ-A-QCDSR-Dr}-\eqref{TetraQ-Negative-V-QCDSR-Dr}. If we only take into account the tetraquark molecular states at the hadron side, we can obtain reasonable molecule masses but  overestimate the pole residues.

\section{Conclusion}

The quarks and gluons are confined objects, they cannot be put on the mass-shell, it is questionable  to use the Landau equation to study  the quark-gluon bound states.
Furthermore,  we carry out the operator product expansion in the deep Euclidean region $p^2\to -\infty$ in the QCD sum rules, where the Landau singularities cannot exist.
If we insist on applying  the Landau equation to study the Feynman diagrams in the QCD sum rules, we should choose the pole masses rather than the $\overline{MS}$ masses,  which lead to obvious problems in the QCD sum rules for the traditional or normal  charmonium and bottomonium states.

 Lucha, Melikhov and Sazdjian assert that the contributions at the order $\mathcal{O}(\alpha_s^k)$ with $k\leq1$ in the operator product expansion, which are factorizable in the color space, are exactly canceled out by the meson-meson scattering states, the nonfactorizable diagrams in the color space, if have a Landau singularity, begin to make contributions  to the tetraquark (molecular) states, the tetraquark molecular states begin to receive contributions  at the order $\mathcal{O}(\alpha_s^2)$.
 In fact, the nonfactorizable Feynman diagrams begin to appear at the order $\mathcal{O}(\alpha_s^0/\alpha_s^1)$ rather than at the order $\mathcal{O}(\alpha_s^2)$, and make contributions  to the tetraquark molecular states. Furthermore, the Landau singularities obtained by Lucha, Melikhov and Sazdjian are questionable, as the Landau singularities appear  at the region $p^2\geq(\hat{m}_{u/s}+\hat{m}_{d/s}+\hat{m}_c+\hat{m}_c)^2$ rather than at the deep Euclidean region $p^2\to -\infty$.

The meson-meson scattering state and tetraquark molecular state both have four valence quarks, which form two color-neutral clusters, we cannot distinguish which Feynman diagrams contribute to the  meson-meson scattering state or tetraquark molecular state based on the two color-neutral clusters in the factorizable Feynman diagrams.
The Landau equation servers  as a kinematical equation in the momentum space, and is independent on the factorizable and nonfactorizable properties of the Feynman diagrams in the color space.

We choose the axialvector current $J_\mu(x)$ and tensor current $J_{\mu\nu}(x)$ to examine the outcome if the assertion of Lucha, Melikhov and Sazdjian
is right.
The  axialvector current $J_\mu(x)$ couples potentially to the  charged $D\bar{D}^*+D^*\bar{D}$ meson-meson scattering states or tetraquark molecular states with the  $J^{PC}=1^{+-}$, while the tensor current $J_{\mu\nu}(x)$ couples potentially to the neutral $D_s^*\bar{D}_{s1}-D_{s1}\bar{D}_s^*$ meson-meson scattering states or tetraquark molecular states with the $J^{PC}=1^{-+}$. The quantum numbers of the $D\bar{D}^*+D^*\bar{D}$ and $D_s^*\bar{D}_{s1}-D_{s1}\bar{D}_s^*$ differ from the traditional or normal  mesons significantly, and  are good subjects to study the exotic states.
After detailed analysis, we observe that the meson-meson scattering states cannot saturate the QCD sum rules,  while
the tetraquark molecular states can saturate the QCD sum rules. We can take into account the meson-meson scattering states reasonably by adding a finite width to the
tetraquark molecular states.

The Landau equation is useless to study the Feynman diagrams in the QCD sum rules for the tetraquark molecular states,  the   tetraquark molecular states begin to receive contributions at the order $\mathcal{O}(\alpha_s^0/\alpha_s^1)$ rather than at the order $\mathcal{O}(\alpha_s^2)$.

\section*{Appendix}

The explicit expressions of the QCD spectral densities $\rho_{Z,QCD}(s)$ and $\rho_{X,QCD}(s)$,
\begin{eqnarray}
\rho_{Z,QCD}(s)&=&\rho^Z_0(s)+\rho^Z_3(s)+\rho^Z_4(s)+\rho^Z_5(s)+\rho^Z_6(s)+\rho^Z_7(s)+\rho^Z_8(s)+\rho^Z_{10}(s)\, ,\nonumber\\
\rho_{X,QCD}(s)&=&\rho^X_0(s)+\rho^X_3(s)+\rho^X_4(s)+\rho^X_5(s)+\rho^X_6(s)+\rho^X_7(s)+\rho^X_8(s)+\rho^X_{10}(s)\, ,
\end{eqnarray}

\begin{eqnarray}
\rho^Z_0(s)&=&\frac{1}{4096\pi^6}\int dydz \, yz\left(1-y-z\right)^3\left(s-\overline{m}_c^2\right)^2
\left(35s^2-26s\overline{m}_c^2+3\overline{m}_c^4\right) \, ,
\end{eqnarray}

\begin{eqnarray}
\rho^Z_3(s)&=&-\frac{3m_c\langle\bar{q}q\rangle}{256\pi^4} \int dydz\, (y+z)\left(1-y-z\right)\left(s-\overline{m}_c^2\right)\left(7s-3\overline{m}_c^2\right)\, ,
\end{eqnarray}

\begin{eqnarray}
\rho^Z_{4}(s)&=&-\frac{m_c^2}{3072\pi^4}\langle\frac{\alpha_{s}GG}{\pi}\rangle \int dydz
\left(\frac{z}{y^2}+\frac{y}{z^2}\right)\left(1-y-z\right)^3\left[8s-3\overline{m}_c^2 +s^2\delta\left(s-\overline{m}_c^2\right)\right]  \nonumber\\
&&+\frac{1}{1024\pi^4}\langle\frac{\alpha_{s}GG}{\pi}\rangle \int dydz
\,(y+z)\left(1-y-z\right)^2 s\left(5s-4\overline{m}_c^2\right)  \, ,
\end{eqnarray}

\begin{eqnarray}
\rho^Z_5(s)&=&\frac{3m_c\langle\bar{q}g_{s}\sigma Gq\rangle}{512\pi^4} \int dydz\, (y+z)\left(5s-3\overline{m}_c^2\right) \nonumber\\
&&-\frac{3m_c\langle\bar{q}g_{s}\sigma Gq\rangle}{256\pi^4} \int dydz  \left(\frac{y}{z}+\frac{z}{y}\right)\left(1-y-z\right)
\left(2s-\overline{m}_c^2\right) \, ,
\end{eqnarray}

\begin{eqnarray}
\rho^Z_6(s)&=&\frac{m_c^2\langle\bar{q}q\rangle^2}{16\pi^2} \int dy  +\frac{g_s^2\langle\bar{q}q\rangle^2}{864\pi^4} \int dydz\, yz
\left[8s-3\overline{m}_c^2+s^2\delta\left(s-\overline{m}_c^2\right)\right]\nonumber\\
&&-\frac{g_s^2\langle\bar{q}q\rangle^2}{576\pi^4} \int dydz
\left(\frac{z}{y}+\frac{y}{z}\right)\left(1-y-z\right) \left(7s-4\overline{m}_c^2\right) \nonumber\\
&&-\frac{m_c^2g_s^2\langle\bar{q}q\rangle^2}{1728\pi^4} \int dydz
\left(\frac{z}{y^2}+\frac{y}{z^2}\right)\left(1-y-z\right)\left[7+5s\,\delta\left(s-\overline{m}_c^2\right)\right]\nonumber\\
&&+\frac{g_s^2\langle\bar{q}q\rangle^2}{1728\pi^4} \int dydz
(y+z)  \left(1-y-z\right) \left(4s-3\overline{m}_c^2\right)\, ,
\end{eqnarray}

\begin{eqnarray}
\rho^Z_{7}(s)&=&\frac{m_c^3\langle\bar{q}q\rangle}{768\pi^2} \langle\frac{\alpha_{s}GG}{\pi}\rangle \int dydz\,
(y+z)\left(\frac{1}{z^3}+\frac{1}{y^3}\right)\left(1-y-z\right)
\left(1+\frac{2s}{T^2}\right)\delta\left(s-\overline{m}_c^2\right)  \nonumber\\
&&-\frac{3m_c\langle\bar{q}q\rangle}{256\pi^2} \langle\frac{\alpha_{s}GG}{\pi}\rangle \int dydz
\left(\frac{y}{z^2}+\frac{z}{y^2}\right)\left(1-y-z\right)
\left[1+\frac{2s}{3}\delta\left(s-\overline{m}_c^2\right)\right]\nonumber\\
&&-\frac{m_c \langle\bar{q}q\rangle}{128\pi^2} \langle\frac{\alpha_{s}GG}{\pi}\rangle \int dy dz
\left[1+\frac{2s}{3}\delta\left(s-\overline{m}_c^2\right)\right]\nonumber\\
&&-\frac{m_c\langle\bar{q}q\rangle}{512\pi^2} \langle\frac{\alpha_{s}GG}{\pi}\rangle \int dy
 \left[1+\frac{2s}{3}\delta\left(s-\widetilde{m}_c^2\right)\right]\, ,
\end{eqnarray}

\begin{eqnarray}
\rho^Z_8(s)&=&-\frac{m_c^2\langle\bar{q}q\rangle \langle\bar{q}g_{s}\sigma Gq\rangle}{32\pi^2}  \int dy\left(1+\frac{s}{T^2}\right)
\delta\left(s-\widetilde{m}_c^2\right)  \nonumber\\
&&+\frac{ \langle\bar{q}q\rangle \langle\bar{q}g_{s}\sigma Gq\rangle}{64\pi^2} \int dy\, s\,
\delta\left(s-\widetilde{m}_c^2\right)\, ,
\end{eqnarray}

\begin{eqnarray}
\rho^Z_{10}(s)&=& -\frac{  \langle\bar{q}g_{s}\sigma Gq\rangle^2}{256\pi^2T^4} \left(1-\frac{m_c^2}{T^2} \right)\int dy \,s^2\,\delta\left(s-\widetilde{m}_c^2\right) \nonumber\\
&&-\frac{m_c^4\langle\bar{q}q\rangle^2}{288T^4} \langle\frac{\alpha_{s}GG}{\pi}\rangle \int dy \left[\frac{1}{y^3}+\frac{1}{\left(1-y\right)^3}\right]
\delta\left(s-\widetilde{m}_c^2\right)  \nonumber\\
&&+\frac{m_c^2\langle\bar{q}q\rangle^2}{96T^2} \langle\frac{\alpha_{s}GG}{\pi}\rangle \int dy
\left[\frac{1}{y^2}+\frac{1}{\left(1-y\right)^2}\right]\delta\left(s-\widetilde{m}_c^2\right)\nonumber\\
&&+\frac{\langle\bar{q}g_{s}\sigma Gq\rangle^2}{18432\pi^2} \int dy\left(1+\frac{2s}{T^2}\right)
\delta\left(s-\widetilde{m}_c^2\right)\nonumber\\
&&+\frac{m_c^2\langle\bar{q}q\rangle^2}{288T^6} \langle\frac{\alpha_{s}GG}{\pi}\rangle \int dy
s^2\delta\left(s-\widetilde{m}_c^2\right) \, ,
\end{eqnarray}

\begin{eqnarray}
\rho^X_{0}(s)&=&\frac{1}{2048\pi^6}\int dydz\, yz\left(1-y-z\right)^2  \left(s-\overline{m}_c^2\right)^3\left(6s-\overline{m}_c^2\right) \nonumber\\
&&+\frac{1}{8192\pi^6}\int dydz\, yz\left(1-y-z\right)^3\left(s-\overline{m}_c^2\right)^2\left(33s^2-18s\overline{m}_c^2+\overline{m}_c^4\right) \, ,
\end{eqnarray}

\begin{eqnarray}
\rho^X_{3}(s)&=&\frac{m_s\langle\bar{s}s\rangle}{128\pi^4}\int dydz\, yz \left(s-\overline{m}_c^2\right) \left(7s-2\overline{m}_c^2\right) \nonumber\\
&&+\frac{m_s\langle\bar{s}s\rangle}{256\pi^4} \int dydz\, yz\left(1-y-z\right)\left(35s^2-30s\overline{m}_c^2+3\overline{m}_c^4\right) \nonumber\\
&&+\frac{9m_s m_c^2\langle\bar{s}s\rangle}{64\pi^4}\int dydz\, \left(s-\overline{m}_c^2\right)\, ,
\end{eqnarray}

\begin{eqnarray}
\rho^X_{4}(s)&=&-\frac{m_c^2}{6144\pi^4}\langle\frac{\alpha_{s}GG}{\pi}\rangle \int dydz
\left(\frac{z}{y^2}+\frac{y}{z^2}\right) \left(1-y-z\right)^2\left(9s-4\overline{m}_c^2\right)  \nonumber\\
&&-\frac{m_c^2}{6144\pi^4}\langle\frac{\alpha_{s}GG}{\pi}\rangle \int dydz \left(\frac{z}{y^2}+\frac{y}{z^2}\right)\left(1-y-z\right)^3
\left[5s-\overline{m}_c^2 +\frac{4s^2}{3}\delta\left(s-\overline{m}_c^2\right)\right]  \nonumber\\
&&-\frac{1}{3072\pi^4}\langle\frac{\alpha_{s}GG}{\pi}\rangle \int dydz \, (y+z) \left(1-y-z\right)
\left(s-\overline{m}_c^2\right)\left(20s-7\overline{m}_c^2\right) \nonumber\\
&&+\frac{1}{12288\pi^4}\langle\frac{\alpha_{s}GG}{\pi}\rangle \int dydz\, (y+z)\left(1-y-z\right)^2
\left(35s^2-30s\overline{m}_c^2+3\overline{m}_c^4\right)\, ,
\end{eqnarray}

\begin{eqnarray}
\rho^X_{5}(s)&=&-\frac{m_s\langle\bar{s}g_{s}\sigma Gs\rangle}{768\pi^4} \int dy\, y\left(1-y\right)\left(9s-4\widetilde{m}_c^2\right)  \nonumber\\
&&-\frac{m_s\langle\bar{s}g_{s}\sigma Gs\rangle}{256\pi^4} \int dydz\, yz \left[5s-\overline{m}_c^2
+\frac{4s^2}{3}\delta\left(s-\overline{m}_c^2\right)\right] \nonumber\\
&&-\frac{9m_s m_c^2\langle\bar{s}g_{s}\sigma Gs\rangle}{256\pi^4} \int dy \, ,
\end{eqnarray}

\begin{eqnarray}
\rho^X_{6}(s)&=&-\frac{3m_c^2\langle\bar{s}s\rangle^2}{32\pi^2} \int dy\, ,
\end{eqnarray}

\begin{eqnarray}
\rho^X_{7}(s)&=&\frac{m_s m_c^2\langle\bar{s}s\rangle}{2304\pi^2} \langle\frac{\alpha_{s}GG}{\pi}\rangle \int dydz
\left(\frac{z}{y^2}+\frac{y}{z^2}\right)\left(1-\frac{5s}{T^2}\right)\delta\left(s-\overline{m}_c^2\right)  \nonumber\\
&&+\frac{m_s m_c^2\langle\bar{s}s\rangle}{576\pi^2} \langle\frac{\alpha_{s}GG}{\pi}\rangle \int dydz\,
\left(\frac{z}{y^2}+\frac{y}{z^2}\right) \left(1-y-z\right)\left(\frac{1}{4}+\frac{s}{T^2}-\frac{s^2}{T^4}\right)
\delta\left(s-\overline{m}_c^2\right)  \nonumber\\
&&+\frac{m_s m_c^2\langle\bar{s}s\rangle}{64\pi^2} \langle\frac{\alpha_{s}GG}{\pi}\rangle \int dydz
 \frac{1}{y^2} \left(2-y -\frac{y s}{T^2}\right)\delta\left(s-\overline{m}_c^2\right)  \nonumber\\
&&-\frac{m_s\langle\bar{s}s\rangle}{4608\pi^2}\langle\frac{\alpha_{s}GG}{\pi}\rangle \int dy
 \left[14+13s\delta\left(s-\widetilde{m}_c^2\right)\right] \nonumber\\
&&+\frac{m_s\langle\bar{s}s\rangle}{1152\pi^2}\langle\frac{\alpha_{s}GG}{\pi}\rangle \int dydz\,
(y+z)\left[\frac{3}{4}+\left(s+\frac{s^2}{T^2}\right)\delta\left(s-\overline{m}_c^2\right)\right] \nonumber\\
&&+\frac{m_s m_c^2\langle\bar{s}s\rangle}{256\pi^2} \langle\frac{\alpha_{s}GG}{\pi}\rangle \int dy \left(
1+\frac{s}{T^2}\right)\delta\left(s-\widetilde{m}_c^2\right) \, ,
\end{eqnarray}

\begin{eqnarray}
\rho^X_{8}(s)&=&\frac{3m_c^2\langle\bar{s}s\rangle \langle\bar{s}g_{s}\sigma Gs\rangle}{64\pi^2} \int dy
\left(1+\frac{s}{T^2}\right)\delta\left(s-\widetilde{m}_c^2\right) \, ,
\end{eqnarray}

\begin{eqnarray}
\rho^X_{10}(s)&=&-\frac{3m_c^2 \langle\bar{s}g_{s}\sigma Gs\rangle^2}{512\pi^2T^6} \int dy\,  s^2 \delta\left(s-\widetilde{m}_c^2\right)\nonumber\\
&&+\frac{m_c^2\langle\bar{s}s\rangle^2}{48T^2} \langle\frac{\alpha_{s}GG}{\pi}\rangle \int dy\, \frac{1}{y^2}
\left(1-\frac{y s}{2T^2}\right)\delta\left(s-\widetilde{m}_c^2\right)\nonumber\\
&&+\frac{\langle\bar{s}g_{s}\sigma Gs\rangle^2}{3072\pi^2T^2}\int dy \, s\, \delta\left(s-\widetilde{m}_c^2\right)\nonumber\\
&&-\frac{m_c^2\langle\bar{s}s\rangle^2}{192T^6} \langle\frac{\alpha_{s}GG}{\pi}\rangle \int dy\,
s^2\delta\left(s-\widetilde{m}_c^2\right)\, ,
\end{eqnarray}
where $\int dydz=\int_{y_i}^{y_f}dy\int_{z_i}^{1-y}dz$, $\int dy=\int_{y_i}^{y_f} dy$, $y_{f}=\frac{1+\sqrt{1-4m_c^2/s}}{2}$,
$y_{i}=\frac{1-\sqrt{1-4m_c^2/s}}{2}$, $z_{i}=\frac{ym_c^2}{y s -m_c^2}$, $\overline{m}_c^2=\frac{(y+z)m_c^2}{yz}$,
$ \widetilde{m}_c^2=\frac{m_c^2}{y(1-y)}$, $\int_{y_i}^{y_f}dy \to \int_{0}^{1}dy$, $\int_{z_i}^{1-y}dz \to \int_{0}^{1-y}dz$ when the $\delta$ functions $\delta\left(s-\overline{m}_c^2\right)$ and $\delta\left(s-\widetilde{m}_c^2\right)$ appear.

\section*{Acknowledgements}
This  work is supported by National Natural Science Foundation, Grant Number  11775079.

\end{document}